\documentclass[]{jfm}

\usepackage{graphicx}
\usepackage{graphbox}
\usepackage{newtxtext}
\usepackage{newtxmath}
\usepackage{mathtools}
\usepackage{natbib}
\usepackage{hyperref}
\hypersetup{
	colorlinks = true,
	urlcolor   = blue,
	citecolor  = black,
}

\usepackage{multirow}
\usepackage{accents}
\usepackage{comment}
\usepackage{subcaption}
\usepackage{xcolor} 
\usepackage{epstopdf, epsfig}
\usepackage{tikz}

\newcommand{\RomanNumeralCaps}[1]
\nolinenumbers
\graphicspath{{figures/}}
\usepackage{cleveref} 

\newcommand{\mbf}[1]{\boldsymbol{#1}}

\shorttitle{Early turbulence in viscoelastic FPCA
}

\shortauthor{L. Zhu and R. R. Kerswell}

\title{Early turbulence in viscoelastic flow past a periodic cylinder array }

\author{Lu Zhu\aff{1} \corresp{\email{lz447@cam.ac.uk}}, 
     Rich R. Kerswell\aff{1}}

 \affiliation{\aff{1}Department of Applied Mathematics and Theoretical Physics, University of Cambridge, Wilberforce Road, Cambridge CB3 0WA, UK}

\begin{document}

\maketitle

\begin{abstract}
Early turbulence in periodic cylinder arrays is of particular interest in many practical applications to enhance mixing and material/heat exchange. In this study, we reveal a new early transition pathway to a chaotic wavy state and drag enhancement with the addition of polymers.  Using 2D direct numerical simulations with sufficiently small polymer diffusion ($\epsilon=10^{-5}$), we show that viscoelastic flow past periodic cylinder arrays become unstable at a Reynolds number $Re\approx 10$, significantly lower than the Newtonian counterpart of $Re\approx 150-200$. The chaotic wavy state which ensues exhibits sheets and `arrowhead' polymer conformation structures, consistent with the saturated centre-mode instability observed in wall-bounded parallel flows (Page et al. {\em Phys. Rev. Lett.} {\bf 125}, 154501, 2020). Analysis of the kinetic energy budget reveals the purely elastic origin of the chaos. However, inertial forces, in conjunction with elastic forces, can reshape the base state, affecting the formation of an invariant polymer sheet. This sheet facilitates the stretching and recoiling of polymers, which in turn induces flow fluctuations and maintains the chaos. Exploring various maximum polymer extensions $b$, and polymer concentrations $\beta$ highlights the role of elastic forces in stretching the upstream separation zone while suppressing the downstream separation zone, resulting in drag enhancement at finite $Re$. Surprisingly, these modifications to the base state can suppress the invariant polymer sheet under large elastic forces (large $b$ or small $\beta$), thereby achieving a stable polymer-modified laminar state. 
\end{abstract}

\begin{keywords}
vicoelastic fluids, direct numerical simulation, early turbulence and transition, elastic instabilities
\end{keywords}

\section{Introduction}\label{sec:intro}

Dissolving polymers into Newtonian fluids introduces elastic effects that can fundamentally alter flow properties, resulting in phenomena such as drag enhancement or reduction, and the initiation or suppression of various instabilities and turbulence within different flow systems~\citep{xi2019turbulent, steinberg2021elastic,datta2022,dubief2023elasto}. These effects are significant in a wide range of industrial applications~\citep{khomami1997stability,alves2021numerical}, making them a subject of considerable academic and practical interest. In this work, we focus on viscoelastic fluid flow past cylinder arrays (FPCA).

As a benchmark for problems involving flows in porous and fibrous media~\citep{sorbie2013polymer,wei2014oil}, heat exchangers~\citep{fisher1996analysis}, and other applications in the oil, food, and electronic industries, the viscoelastic flow past cylinders has been extensively studied, covering a range from inertialess creeping flow to high-Reynolds-number, inertia-dominated flows~\citep{richter2010simulations,steinberg2021elastic,alves2021numerical}. The majority of the work, however, is in the inertialess creeping flow limit.  
Here, the elastic force significantly alters the flow base, leading to drag enhancement compared to the corresponding Newtonian flows~\citep{khomami1997stability,varshney2018drag}. This base state modification is often associated with the formation and stretching of the upstream separation zone, as demonstrated by recent experimental and numerical studies~\citep{kenney2013large,qin2019flow}. More interestingly, as the elasticity becomes higher, elastic instabilities can be triggered which soon generate a chaotic state termed `elastic turbulence'~\citep{groisman2000elastic} since inertia is absent. Despite observations of elastic turbulence in various geometries~\citep{steinberg2021elastic}, the underlying mechanism for this chaotic state is not clear and may vary across different geometries. 
In FPCA, \citet{mckinley1993wake} conducted experiments on unbounded and unidirectionally bounded  cylinders, identifying wake instabilities downstream of the cylinder, characterized by temporally varying streamlines. 
Further experimental investigation revealed a bistable symmetric breaking and further transition to a time-dependent state~\citep{hopkins2020purely}, which was hypothetically attributed to the wake instabilities~\citep{haward2021bifurcations}.
Alternatively, experiments and simulations at high elasticity numbers ($\mathrm{El}$), defined as the ratio between elastic forces and inertial forces, demonstrated that the growth of the upstream separation and vortex pair can ultimately lose the stability, leading to an unstable chaotic state~\citep{kenney2013large,qin2019flow}.

At finite $Re$, viscoelastic flows are known to suppress inertial turbulence by suppressing turbulent intensity~\citep{richter2010simulations} and causing significant drag reduction (DR) in shear flows. Since the first discovery by~\citet{Toms_P1INTCRHEOL1948}, many studies have been conducted to understand this phenomenon but are mostly focussed on wall-bounded parallel flows~\citep{sureshkumar1997direct,kim2008dynamics,white2012re,xi2012intermittent}. In unbounded FPCA, early experiments~\citep{gadd1966effects} have shown polymers can decrease vortex shedding frequency.
Additional studies \citep{azaiez1994numerical,cadot2000experimental,oliveira2001method,richter2010simulations} also demonstrated that polymers effectively suppress wake instabilities and vortical structures, and extend the downstream separation zone, thereby reducing friction drag. This stabilizing effect is expected to postpone the onset of turbulent transition, a finding supported by \citet{cadot2000experimental}.
\citet{richter2010simulations} linked these alterations to drag reduction (DR) in parallel shear flows, where inertial forces dominate and elastic forces contribute to dissipating turbulent kinetic energy.  
Interestingly, with somewhat contradictory results, \citet{james1971drag} reported an enhancement of drag at $Re<200$ in their experiments. Mapping across the Reynolds number ($Re$)-Weissenberg number ($Wi$) parameter space, \citet{xiong2017numerical} show that the viscoelastic flows can exhibit drag enhancement at a finite $Re$, resembling those observed in inertialess flows. 
In these studies, the drag enhancement is attributed to the modification of the laminar base flow, while polymers are still believed to promote flow stability.

Despite the conventional understanding that polymers primarily suppress instabilities at finite $Re$, a new type of turbulence with distinctly different characteristics and driving mechanisms has recently been identified in parallel flow geometries at $Re\sim O(1000)$~\citep{samanta2013elasto,dubief2013mechanism}. This so-called `elasto-inertial turbulence' (EIT) is characterized by sheet-like polymer stress~\citep{sid2018two,Zhu_XiPRFluids2021} and is thought to arise from the combined influences of inertial and elastic effects. Numerical and experimental studies~\citep{samanta2013elasto, sid2018two, choueiri2021experimental, dubief2022} show that EIT can sustain at significantly lower $Re$ compared with Newtonian fluids, thereby promoting an early onset of turbulence. 
The dynamic origin of this early turbulence phenomenon has not yet been conclusively determined. Recent linear stability analysis has revealed a linear unstable centre-mode in viscoelastic pipe flow \citep{garg2018viscoelastic} and channel flow \citep{Khalid2021a} which can exist down to the inertialess limit at least in channel flow \citep{khalid2021continuous}  suggesting the purely elastic nature of the instability \citep{Buza2022a}. This instability, however, exists at higher Weisssenberg numbers than those typical for EIT.  The nonlinear evolution of the centre-mode instability leads to the formation of saturated 'arrowhead' travelling waves, characterized by  polymer stress sheets stretching from either wall to meet forming an arrowhead at the center of the channel~\citep{page2020exact, buza2022finite, morozov2022coherent}.  Fascinatingly, these arrowheads reach down to much lower Weissenberg numbers but seem to coexist with EIT rather than becoming unstable to produce EIT \citep{beneitez2024multistability} . 
Alternatively, \citet{shekar2019critical, shekar2021} have linked EIT to nonlinear viscoelastic Tollmien–Schlichting waves which connect to Newtonian Tollmein-Schlichting waves at least at high $Re$ ($\gtrsim 10,000$). Finally, recent studies~\citep{zhu2022direct,beneitez2023,Couchman2024} have identified linearly unstable wall modes, which may also be responsible for the formation of EIT. 

While there has been a lot work studying EIT in parallel flow geometries, the presence of such a chaotic state in flows with curved streamlines has not been reported before. It is of particular interest to explore the potential connection between EIT and ET, which is well studied in  curved streamline geometries where hoop stress instabilities are operative \citep{shaqfeh1996purely}. This motivated us to study viscoelastic FPCA over a wide range of $Re$. From a practical standpoint, studying viscoelastic flow past cylinders at finite $Re$ has significant importance due to its potential to enhance mixing and heat exchange rates in heat exchangers. While many studies focus on unbounded viscoelastic FPCA \citep{richter2010simulations} or bounded FPCA between parallel walls \citep{grilli2013transition,xiong2017numerical}, we employ periodic cylinders as adopted by \citet{talwar1995flow} to mimick the actual flow patterns observed in pin heat exchangers and porous media.

In this study, we perform a series of 2D simulations of viscoelastic flow past periodic arrays of cylinders to explore the interesting early transition to turbulence and its potential relation to EIT/ET. The numerical model formulation is detailed in \S\ref{sec:method}. The early transition phenomenon and drag enhancement findings are presented in \S\ref{sec:results}. The physical origin of the early wavy and turbulent structures is discussed in \S\ref{sec:nature}.

%
%
\section{Methodology}\label{sec:method}
\subsection{Problems}\label{sec:problem}

\begin{figure}
    \centering    
    \includegraphics[width=0.3\linewidth, trim=0mm 0mm 0mm 0mm, clip]{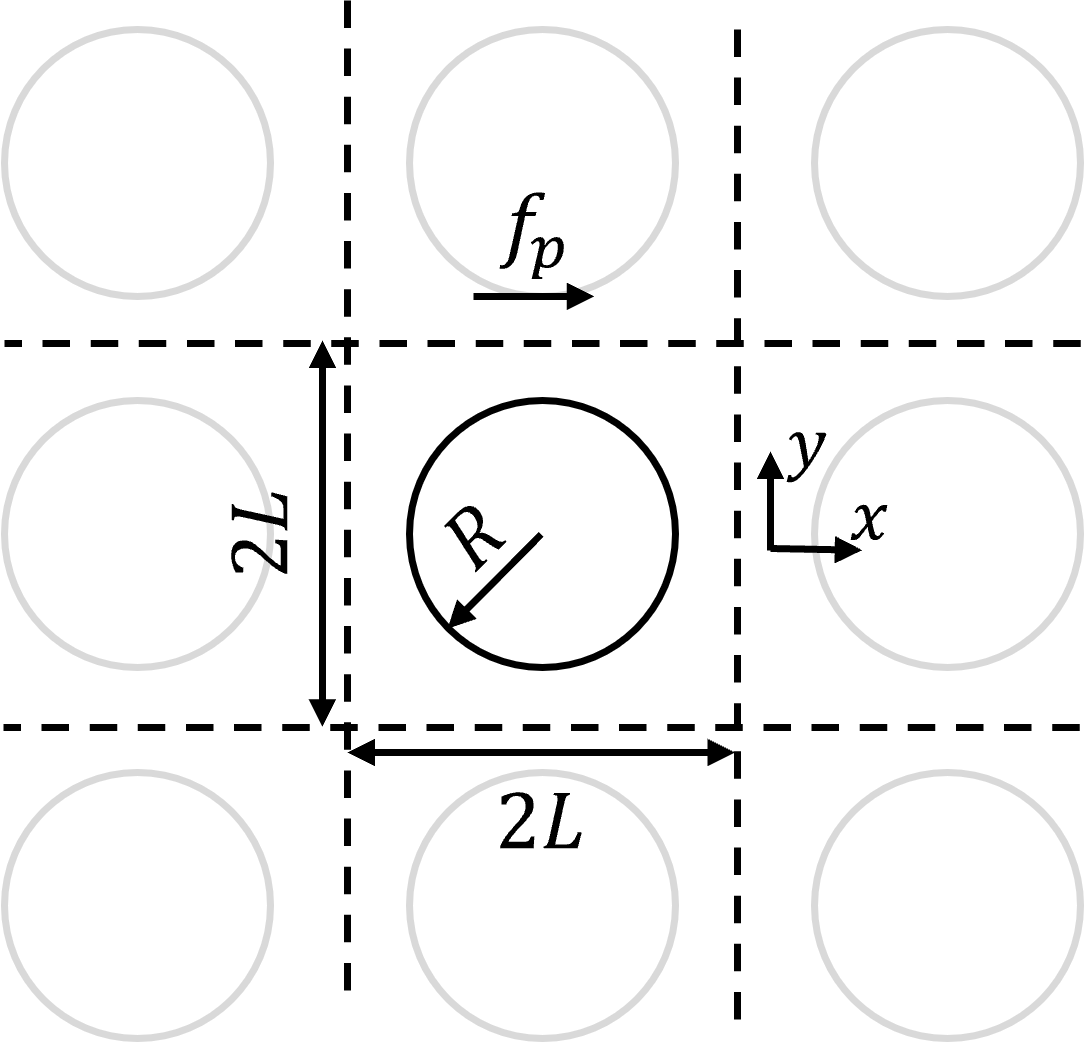}
    \caption{The computational cell is a $2L$ $\times$ $2L$ square at the centre of which is positioned an impenetrable cylinder of radius $R$. Periodic boundary conditions are imposed across the cell walls.}
    \label{fig:geom}
\end{figure}

A square $2L \times 2L$ computational cell is adopted to model flow in periodic cylinder arrays. The configuration of the computational domain is shown in \cref{fig:geom} where a cylinder of radius $R=L/2$ is placed at the cell centre. The flow is driven by a constant pressure drop $f_p/Re$ pointing in the  $x$-direction. Periodic boundary conditions are applied across both sets of boundaries of the cell. This configuration has been both experimentally and numerically adopted by~\citet{talwar1995flow} and \citet{khomami1997stability}. The dimensional volumetric flow rate of the benchmark Newtonian laminar flow, $Q_{\text{Newt}}$, and the half-cell length $L$ are used to nondimensionalise the system so that, for example, velocities are scaled by the characteristic speed $U=Q_{\text{Newt}}/2L$ and times by $L/U$. 

The viscoelastic flow in this system is described by the continuity, momentum and FENE-P (Finitely Extensible
Nonlinear Elastic model with the Peterlin approximation) constitutive equations~\citep{bird1987dynamics}:

\begin{eqnarray}
\label{eq:ns:mom}%
\frac{D \mbf{u}}{D t} =
- \mbf{\nabla}p + \frac{\beta}{\mathrm{Re}} \nabla^{2}\mbf{u} 
+ \frac{1 -\beta}{\mathrm{Re}}\left(\mbf{\nabla} \cdot
\mbf{\tau}_p\right)+\frac{f_p}{\Rey}-\frac{\Gamma(\boldsymbol{x})}{\varsigma}\mbf{u},%
\\
\label{eq:fenep:conf}
\frac{D\mbf{\alpha}}{D t} 
-
\mbf{\alpha} \cdot \mbf{\nabla u} - \left( \mbf{\alpha} \cdot \mbf{\nabla u}
\right)^{\mathrm{T}} - \epsilon\nabla^{2}\mbf{\mbf{\alpha}} 
= -\mbf{\tau}_p -\frac{\Gamma(\boldsymbol{x})}{\varsigma}(\mbf{\alpha}-\mbf{\delta}),
\\
\label{eq:ns:mass}%
\mbf{\nabla} \cdot \mbf{u} = 0,%
\\
\label{eq:fenep:stress}%
\mbf{\tau}_p = \frac{1}{\mathrm{Wi}} \left(\frac{\mbf{\alpha}}{1 -
	\frac{\mathrm{tr}(\mbf{\alpha})-2}{b}} -\mbf{\delta}\right) 
\end{eqnarray}
where $\mbf{\alpha}$ and $\mbf\tau_\text{p}$ are polymer conformation and stress tensors, respectively. $\mathrm{Re}\equiv\rho UL/\eta$ is the Reynolds number,
$\mathrm{Wi}\equiv \lambda U/L$ is the Weissenberg number (where $\lambda$ is the polymer relaxation time), $\beta\equiv\eta_\text{s}/\eta$ is the viscosity ratio between the fluid viscosity $\eta$ and solvent viscosity $\eta_\text{s}$), and $b\equiv\max(\mathrm{tr}\,\mbf{\alpha})$ is the finite extensibility parameter. The forcing term $f_p/Re$ is chosen so that the volumetric flux $Q_\text{Newt}=2L$ giving the corresponding characteristic velocity $U=1$. The last term on the left-hand side of \eqref{eq:fenep:conf} accounts for the polymer stress diffusion (where $\epsilon\equiv 1/\mathrm{(ReSc)}$ is the polymeric diffusion coefficient; $\mathrm{Sc}$ is the Schmidt number) which is often very small compared with the viscous diffusion, i.e. $\mathrm{Sc} \gg 1$.

In this study, a volume penalty method~\citep{angot1999analysis,schneider2015immersed,hester2021improving} is used to model the cylinder in a pseudospectral solver. The last terms on the right-hand side of \eqref{eq:ns:mom} and \eqref{eq:fenep:conf} are the volume penalty terms where 
\begin{equation}
    \Gamma(x,y)\equiv\frac{1}{2}(1+\tanh{\frac{2}{\Delta}(\sqrt{x^2+y^2}-\frac{1}{2})})
\end{equation}
is a smooth mask function for a cylinder located at the unit cell center ($(x,y)=(0,0)$) ($\Delta$ is a damping length scale); $\varsigma$ is a damping time scale. \citet{hester2021improving} shows that an appropriate selection of the damping parameters can improve the convergence of the volume penalty method to second-order numerical accuracy. In this paper, we set $\Delta=2\Delta_x$ ($\Delta_x$ is the grid spacing along $x$-direction) and $\varsigma=\mathrm{Re}(\Delta/2.64823)^2$ adopting \citet{hester2021improving}'s optimal mask profile with the second-order accuracy. An investigation shows a good agreement between the volume penalty method and the standard finite element solver.

\subsection{Numerical set-up}\label{sec:numercs}

The 2D direct numerical simulations (DNS) are performed using an open-source solver, Dedalus~\citep{burns2020dedalus}, which has been utilised and validated in viscoelastic flows as well as a wide range of fluid problems~\citep{lecoanet2016validated,zhu2024LSA,beneitez2024multistability, lellep2024}. The spatial discretisation and time stepping use a Fourier
pseudospectral scheme and a third-order, four-stage, diagonally
implicit–explicit Runge–Kutta scheme~\citep{ascher1997implicit} respectively. Details of the numerical settings for the 2D simulations in this work are listed in \cref{tab:simul_overview}. 
The grid resolution is set to $n_x\times n_y=288\times 288$ for most of the STD, HD, and VD cases, $576\times 576$ for cases with a higher $b=6400$ (LE), lower $\beta=0.7$ (VC) or a lower $\epsilon=0.2\times 10^{-5}$ (VD) parameter values and, maximally, $1152 \times 1152$ to confirm the presence of an invariant polymer sheet in fig \ref{fig:base_profile}. 
A resolution validation was conducted consisting of a series of simulations covering a range of $\mathrm{Re}$ (10 and 100) and $\mathrm{Wi}$ (0 to 15) at $b=1000$, $\beta=0.9$, and $\epsilon=10^{-5}$ as well as runs at a few different $b$, $\beta$, and $\epsilon$ which demonstrated that increasing the resolution does not alter the results and conclusions.

\begin{table}
    \begin{center}
    \def~{\hphantom{0}}
	{
	\begin{tabular}{lcccccc}
		$\textrm{Case}$  & $\mathrm{Re}$  & $\mathrm{Wi}$  & $\epsilon$  & $b$ & $\beta$ \\[3pt]
		\hline
            STD & 0.1, 1, 3, 10, 30, 100 & 1, 2, 3, 5, 8, 9, 10, 15 & $10^{-5}$ & 1000 & 0.9\\[3pt]
            HD & 10, 30, 100 & 1, 2, 3, 5, 10, 15 & $\mathbf{10^{-4}}$ & 1000 & 0.9\\[3pt]
            VD & 10 & 10 & $\mathbf{(10,5,2,1,0.5,0.2)\times 10^{-5}}$ & 1000 & 0.9\\[3pt]
            LE & 10 & 1, 5, 10, 20, 30 & $10^{-5}$ & \textbf{6400} & 0.9\\[3pt]
            VC & 10 & 1, 5, 10, 20, 30 & $10^{-5}$ & 1000 & \textbf{0.7, 0.97}\\[3pt]
            \hline    
	\end{tabular}
	\caption{Summary of the DNS parameter settings. Bold font highlights the distinct parameters changed in these cases. Here, `STD', `HD', `VD', `LE', and `VC' stand for standard, high diffusion, varying diffusion, large extension, and varying concentration cases, respectively. }\label{tab:simul_overview}
	}
	\end{center}
\end{table}

\subsection{Impact of polymer diffusion}\label{sec:polydiff}

\begin{figure}
    \centering    
    \includegraphics[width=0.9\linewidth, trim=0mm 0mm 0mm 0mm, clip]{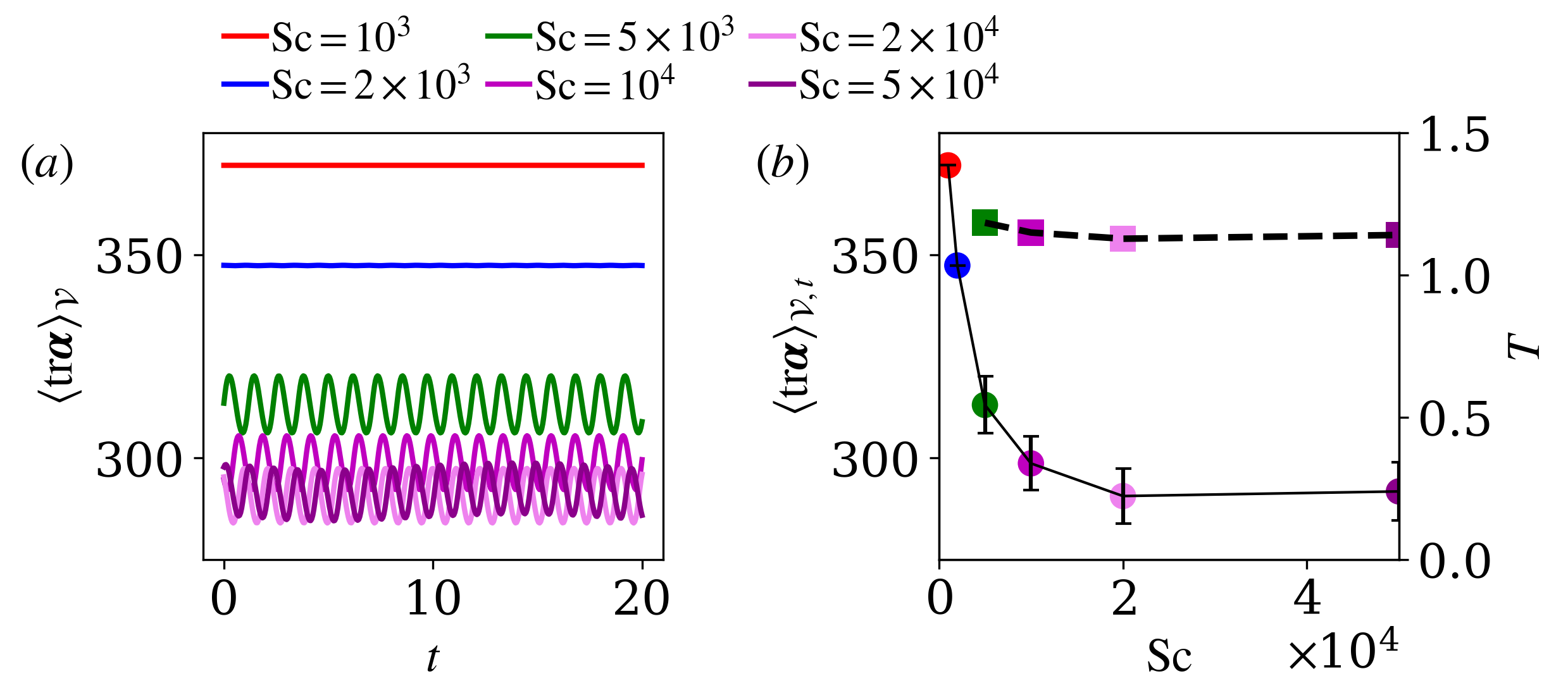}
       
    \caption{Influence of polymer diffusivity $\epsilon$ (VD cases at fixed $(Re,Wi)=(10,10)$): (a) time series of volume averaged polymer conformation trace $\langle\mathrm{tr}\mbf{\alpha}\rangle_{\mathcal{V}}$, (b) dependence of ensemble-averaged $\langle\mathrm{tr}\mbf{\alpha}\rangle_{\mathcal{V},t}$ (circles \& solid line) and oscillation period (squares \& dash line) on $\mathrm{Sc}$.}
    \label{fig:diffusion}
\end{figure}

The potential impact of polymer diffusion on the dynamics of polymeric flows has recently gained significant attention amongst the community~\citep{dubief2023elasto}. The diffusion of the polymer conformation tensor (of the order $10^{-12} \mathrm{m^2/s}$ and, correspondingly, $\mathrm{Sc}=10^6$~\citep{layec1983instability}) stems from the kinetic theory of a dilute polymer solution~\citep{beris1994compatibility} and, historically, since it is  significantly smaller than viscous diffusion, was dropped when formulating the polymer conformation tensor equations~\citep{bird1987dynamics}. However, solving the numerical system without polymer diffusion can introduce numerical difficulty due to the hyperbolic nature of the constitutive equations. Later, polymer diffusion was reinstated into the constitutive equation by~\citet{sureshkumar1997direct} to relieve this numerical difficulty but at highly elevated values so that $\mathrm{Sc}=O(1)$~\citep{housiadas2003polymer,kim2007effects,li2015simple,xi2010turbulent}. Although this choice of numerical diffusion may not obviously affect the simulation of viscoelastic flows in inertia-dominated regimes \citep{Zhu_XiJNNFM2020}, increasing evidence \citep{sid2018two} now suggests the potential risk of suppressing elastic-dominated features of the flow under conditions of high numerical diffusion.

In \cref{fig:diffusion}, we examine the impact of polymer diffusion on the dynamics of the viscoelastic flows. A series of cases (VD in \cref{tab:simul_overview}) with $\mathrm{Sc}$ ranging from $10^3$ to $5\times 10^4$ ($\epsilon$ from $10^{-4}$ to $2\times 10^{-6}$) are compared in terms of the polymer conformation trace, $\mathrm{tr}\mbf{\alpha}$. The chosen $\mathrm{Sc}$ values are significantly larger than $O(1)$ and are comparable to the diffusion rates of realistic polymer solutions~\citep{morozov2022coherent}. As the $\mathrm{Sc}$ increases (polymer diffusivity $\epsilon$ decreases), the average $\langle \mathrm{tr}\mbf{\alpha}\rangle_{\mathcal{V},t}$ (where $\langle \cdot \rangle_{\mathcal{V},t}$ indicates the volumetric and temporal average over the fluid domain $\mathcal{V}$ and time period $\mathcal{T}$, see also \eqref{eq:tram}) diminishes and progressively converges to the value of 290. Notably, as $\mathrm{Sc}$ increases beyond $5\times 10^3$, the flow becomes unstable and exhibits sustained periodic temporal oscillation of $\langle \mathrm{tr}\alpha\rangle_{\mathcal{V}}$ (panel a; $\langle \mathcal{tr}\cdot\rangle_{\mathcal{V}}$ denote volumetric average over $\mathcal{V}$). Once manifest, the wave period $T$ and magnitude become roughly independent of $\mathrm{Sc}$. 
The differing observations under various Schmidt numbers ($Sc$) and corresponding diffusivities ($\epsilon$) suggest the necessity of a sufficiently small $\epsilon$ to capture key phenomena in viscoelastic flows. This $\epsilon$ dependence further implies the elastic nature of these unstable states. 

In the remainder of this paper, we will focus on unveiling these interesting unstable states that have not been previously identified due to the more realistic Sc values used here.

\section{Results: early transition and drag enhancement }\label{sec:results}

In this work, we numerically explore the dynamics of viscoelastic flows at a range of moderate $Re$ in periodic cylinder arrays. Polymer-induced early instabilities, transitions to chaotic turbulence, and drag enhancement are observed and will be discussed in detail.

\subsection{Flow statistics}\label{sec:stat}

To reveal the multiple transition regimes in periodic cylinder arrays, we compute the volumetric flow rate $Q$, the averaged trace of polymer conformation tensor $\langle \mathrm{tr}\mbf{\alpha} \rangle_{\mathcal{V},t}$, and turbulent kinetic energy $\langle E^\prime \rangle_{\mathcal{V},t}$, respectively defined as
\begin{eqnarray}
    \label{eq:Q}
    Q\equiv\frac{1}{2\mathcal{T}} \int_\mathcal{V} \int_\mathcal{T} u \, dt \, dV,
    \\
    \label{eq:tram}
    \langle \mathrm{tr}\mbf{\alpha} \rangle_{\mathcal{V},t}\equiv\frac{1}{\mathcal{V}\mathcal{T}} \int_\mathcal{V} \int_\mathcal{T} \mathrm{tr}\mbf{\alpha} \, dt \, dV,
    \\
    \label{eq:TKE}
    \langle E^\prime \rangle_{\mathcal{V},t}\equiv\frac{1}{\mathcal{V}\mathcal{T}} \int_\mathcal{V} \int_\mathcal{T} \frac{1}{2}(u^{\prime 2} + v^{\prime 2} ) \, dt \, dV
\end{eqnarray}
where $u^\prime = u-\langle u \rangle$ and $v^\prime = v-\langle v \rangle$ (`$\prime$' indicates the fluctuating components of variables; $\langle \cdot \rangle$ denotes average over time) are the velocity fluctuation components along the $x$- and $y$-direction, respectively.  $\mathcal{V}$ and $\mathcal{T}$ denote the fluid domain and the time period encompassing a steady state for statistical integration.

\begin{figure}
      \centering    
      \includegraphics[width=0.98\linewidth, trim=0mm 0mm 0mm 0mm, clip]{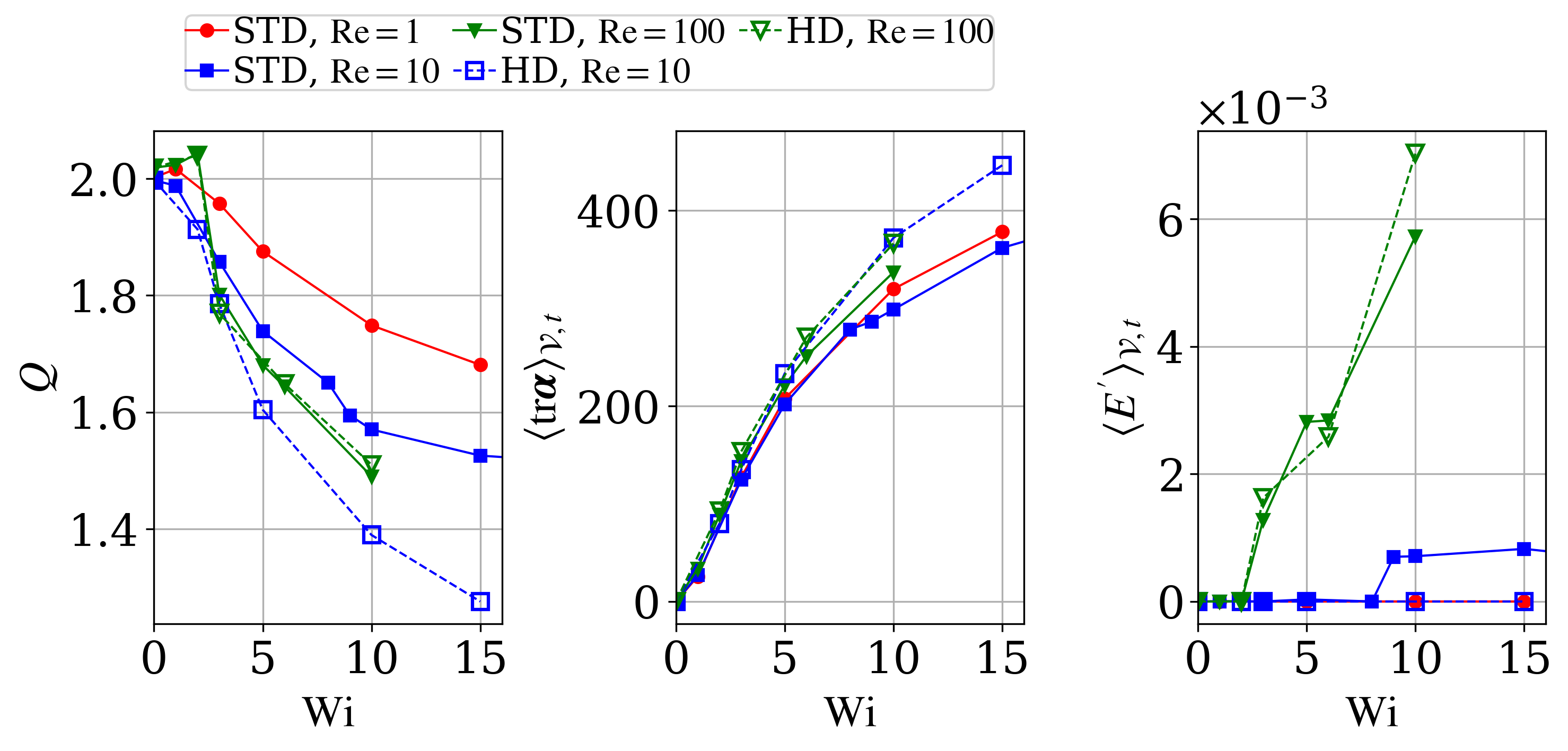}
    \caption{Flow statistics: (a) volumetric flux, (b) trace of polymer conformation tensor, (c) turbulent kinetic energy.}
    \label{fig:stat}
\end{figure}

In \cref{fig:stat}, we show $Q$, $\langle \mathrm{tr}\mbf{\alpha} \rangle_{\mathcal{V},t}$, and $\langle E^\prime \rangle_{\mathcal{V},t}$ of the STD and HD cases at various $Wi$ and $Re$. Due to the geometric confinement between the cylinders in the periodic cylinder arrays, Newtonian flows remain stable until $Re = 150$, as this configuration suppresses further wake development. After polymers are introduced, the polymer extension, measured by $\langle \mathrm{tr}\mbf{\alpha} \rangle_{\mathcal{V},t}$, initially increases nearly linearly with $Wi$ (\cref{fig:stat}(b)). The trend continues until $Wi\approx 5$ after which the nonlinear effects of FENE-P polymer dumbells begin to set in, resulting in a decreased growth rate of profiles. 

In STD cases at $Re=10$, the initial increase in $Wi$ does not significantly change the mean flow statistics, and the volumetric flow rate $Q$ differs by no more than $5\%$ compared to the benchmark Newtonian case. As $Wi>2$, $Q$ starts to decrease with increasing $Wi$, indicating drag enhancement behaviour. This phenomenon has been documented in the literature across both inertialess ($Re\rightarrow 0$) and moderate $Re$ regimes~\citep{james1971drag,talwar1995flow,richter2010simulations,xiong2017numerical,steinberg2021elastic}.  While the drag continuously increases, the turbulent kinetic energy $\langle E^\prime \rangle_{\mathcal{V},t}$ remains zero before $Wi=8$, indicating a steady flow state where polymers only modify the mean flow. From $Wi=8$ to $9$, the TKE profile shoots up to $10^{-3}$ and remains nearly unchanged thereafter, indicating a significantly early transition to an unstable state in viscoelastic flows ($Re\approx 10$) compared with the Newtonian counterpart ($Re>150$) in FPCA. This is, to the knowledge of the authors, the first observation of such an early unsteady state in cylinder arrays at moderate $Re$. The fluctuations of the unsteady state can potentially promote mixing and materials/heat exchange in industrial facilities such as heat exchangers.

The critical $Wi$ for the unstable state depends on $Re$. At $Re=100$, the unstable state appears at $Wi=3$, preceding the polymer-modified laminar stage. Conversely, the chaotic state is not observed at $Re=1$ for $Wi \leq 15$ with the flow remaining in a polymer-modified laminar state as shown in fig. \ref{fig:stat}(a,c). This $Re$-dependence is more clearly seen in \S\ref{sec:regime} and fig. \ref{fig:regime}.  

Another important factor influencing the regime transition is polymer diffusion. As detailed in \S\ref{sec:polydiff}, significant polymer diffusion has the potential to suppress flow oscillations. In \cref{fig:stat}, we show statistics of HD cases with polymer diffusion, $\epsilon$, 10 times larger than the STD cases. At the higher diffusion, we do not observe an unstable state at $Re=10$, highlighting the critical role of low diffusion in triggering these instabilities. Essentially, the polymer diffusion acts to smooth the field of polymer stress which, in return, eliminates small-scale elastic structures and suppresses instabilities. Alternatively, the flow state at $Re=100$ shows relative insensitivity to the polymer diffusion. The profiles of $Q$, $\langle \mathrm{tr}\mbf{\alpha} \rangle_{\mathcal{V},t}$, and $\langle E^\prime \rangle_{\mathcal{V},t}$ collapse well between STD and HD cases. This is hypothesised to be due to the dominance of large-scale elastic streaks in the chaotic state, which are not strongly affected by polymer diffusion. Evidence of these large-scale streaks can be found in \cref{fig:ins_flow}.

The early transition observed here clearly deviates from the common understanding that polymers at moderate $Re$ will suppress the wake instabilities~\citep{richter2010simulations,xiong2017numerical}, thereby delaying their onset. Besides the relatively large polymer diffusion considered in those studies, which might suppress these instabilities, we acknowledge that different geometrical settings could also influence these phenomena. The confinement and periodicity of our geometry likely suppress wake instabilities while potentially promoting others through periodic excitation of the flow. Further investigation into the geometric impact is beyond the scope of this current paper and will be addressed in future studies.

%
%
\subsection{Regime diagrams}\label{sec:regime}

\begin{figure}
    \centering    
    \includegraphics[width=0.7\linewidth, trim=0mm 0mm 0mm 0mm, clip]{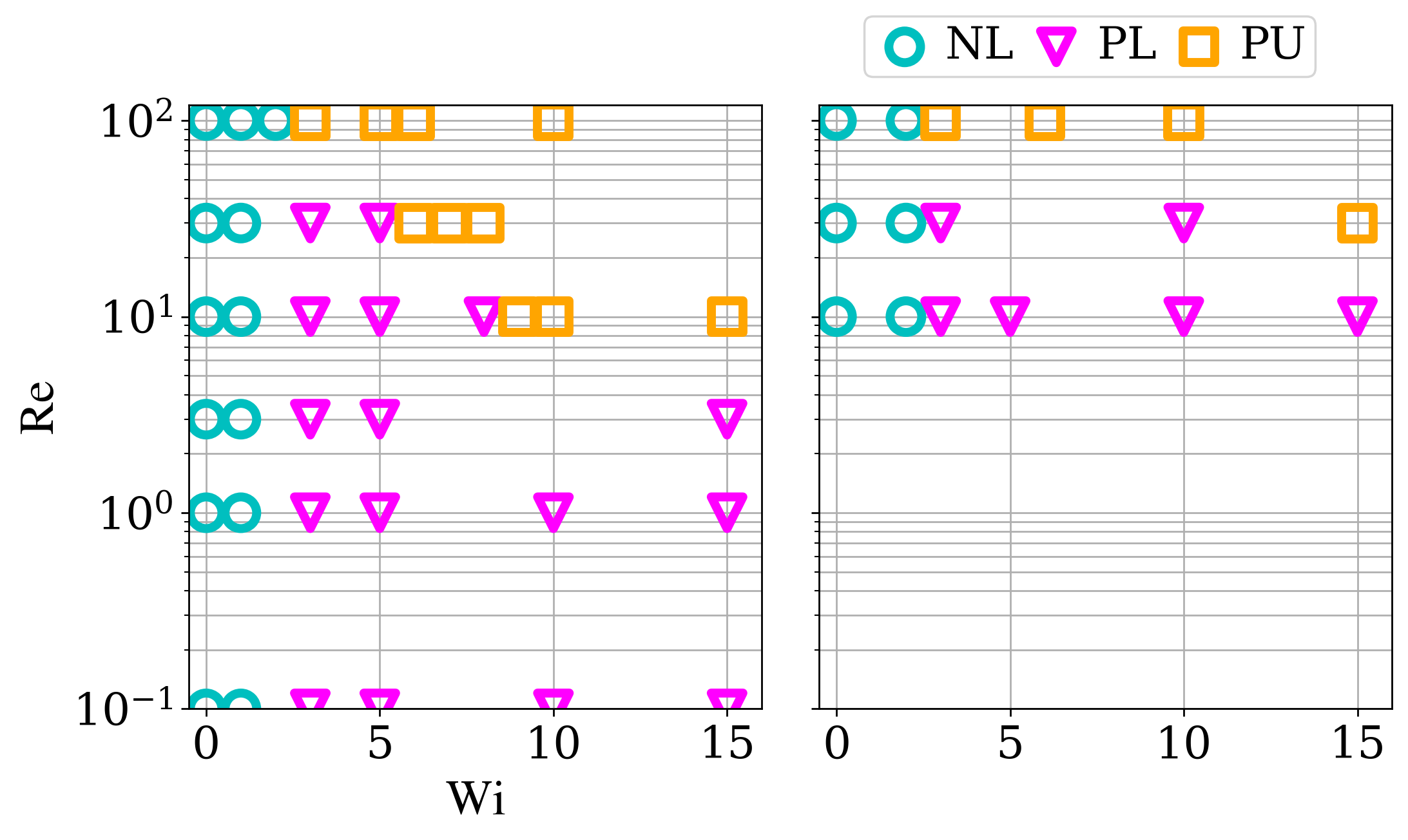}
    \caption{Regime transition in $Re$-$Wi$ parameter space: (a) STD and (b) HD cases. `NL' means Newtonian laminar state, `PL' polymer-modified laminar state and `PU' polymer-modified `unstable' time-dependent state.}
    \label{fig:regime}
\end{figure}

In \cref{fig:regime}, we show the regime transition of STD and HD cases in the $Re$-$Wi$ parameter space. The three distinct states discussed in \S\ref{sec:stat} are henceforth termed Newtonian laminar (NL), polymer-modified laminar (PL), and polymer-modified unstable (PU) states. 
In \cref{fig:regime}, the transition from NL to PL remains independent of $Re$ at $Wi \approx 2$ across 3 orders of magnitude of $Re$ from $10^{-1}$ to $10^2$. This indicates the purely elastic nature of the NL-PL transition, as it is minimally influenced by variations in inertial forces across different $Re$. Moreover, the NL-PL transition is independent of polymer diffusion $\epsilon$, as shown in \cref{fig:regime}(b). This independence arises because the transition modifies the base state, whereas polymer diffusion primarily impacts the small-scale elastic structures resulting from the sharp gradient of polymer stress. 

The onset of PU state, conversely, is influenced by both $Re$ and $Wi$. A higher $Wi$ is required to induce the PU state at lower $Re$. Intuitively, the dependence of the onset of the unstable state on $Re$ may be attributed to the variations in inertial forces that affect flow instabilities, suggesting that the unstable state is elasto-inertial. However, it is important to note that in periodic cylinder arrays, the inertial forces can influence the base state, which in turn influences the formation of instabilities. In this context, the instabilities can be purely elastic. We will investigate the mechanism of PU further in \S\ref{sec:nature}.
Also, note that $\epsilon$ can influence the PL-PU transition. As small-scale structures are suppressed by diffusion, the PU regime at $Re=30$ shifts to higher $Wi$ while absent at $Re=10$.

%
%
\subsection{Flow structures}\label{sec:flowfield}

\begin{figure}
    \centering    
    \includegraphics[width=0.98\linewidth, trim=0mm 0mm 0mm 0mm, clip]{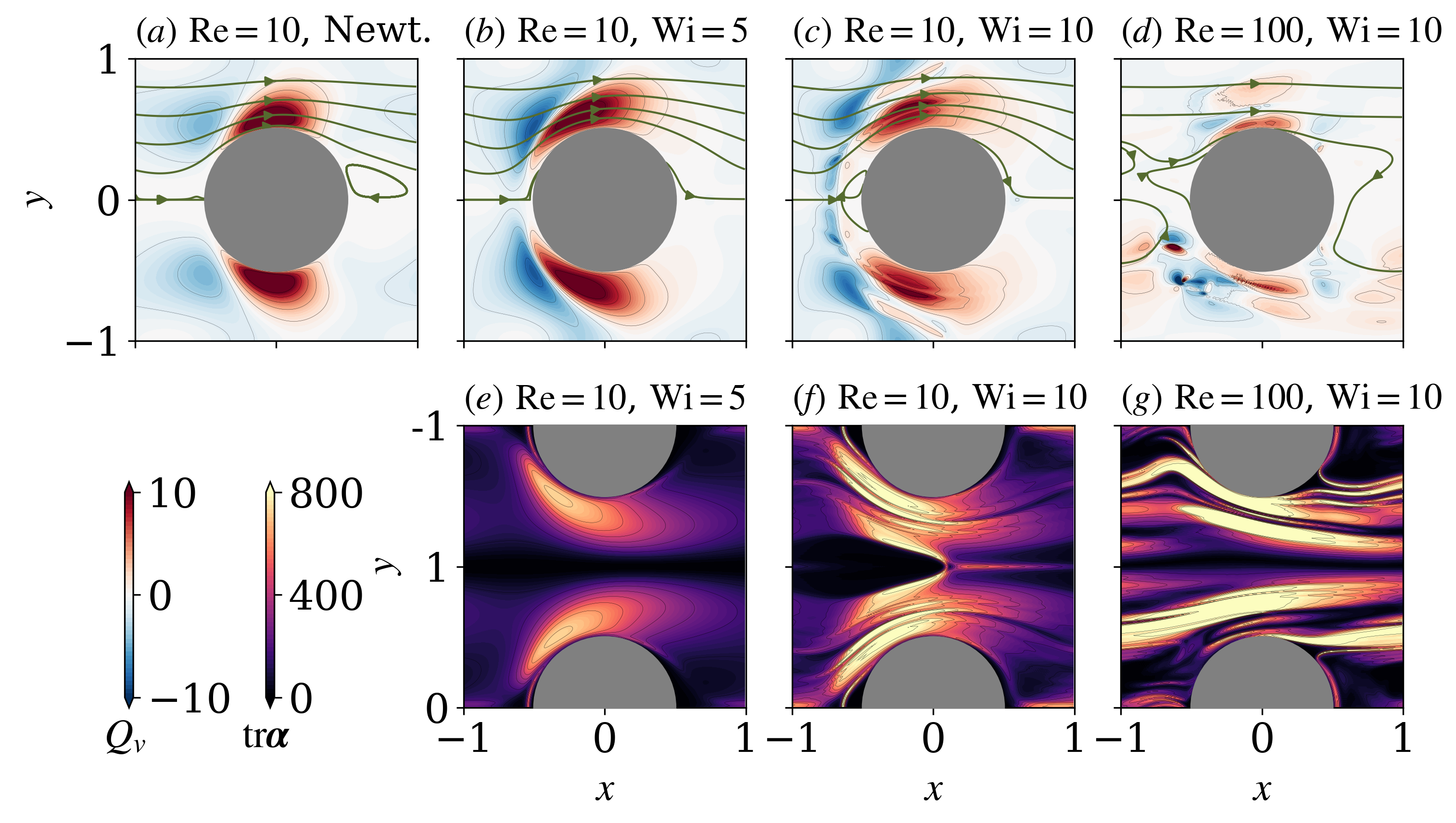}
    \caption{Instantaneous snapshots of STD cases: (a-d) vortices identified by $Q$-criterion, (e-g) trace of polymer conformation tensor. Green lines represent the streamlines of the flow. The $\mathrm{tr}\,\mbf{\alpha}$ field is vertically rolled by 1 to display the complete channel between two cylinders. }
    \label{fig:ins_flow}
\end{figure}

We now examine the instantaneous flow fields in each regime of periodic cylinder arrays. \Cref{fig:ins_flow} shows the instantaneous vortical and polymer extension ($\mathrm{tr}\,\mbf{\alpha})$ fields of four typical STD cases: (i) Newtonian laminar at $Re=10$, (ii) PL at $(Re,Wi)=(10,5)$, (iii) wavy PU at $(Re,Wi)=(10,10)$, and (iv) chaotic PU at $(Re,Wi)=(100,10)$. The $Q$-criterion is used to identify the 2D rotation and stretching motions which is  defined as:
\begin{equation}
    \label{eq:Q}
    Q_v\equiv\frac{1}{2}(\vert \boldsymbol{\Omega} \vert^2 - \vert \boldsymbol{S} \vert^2 )=-\frac{1}{2} 
    \biggl[
    \left(\frac{\partial u}{\partial x}\right)^2 + \left(\frac{\partial v}{\partial y}\right)^2 
    \biggr]-\frac{\partial u}{\partial y}\frac{\partial v}{\partial x},
\end{equation}
where $\boldsymbol{\Omega}$ and $\boldsymbol{S}$ are the rotation rate and strain rate tensors.

%
%

In \cref{fig:ins_flow}(a-d), the presence of the cylinders creates a contracted channel between them.  Flow passing through this contraction therefore has curved streamlines, identified as large-scale rotational motion (red regions on the upper and lower sides of the cylinder) by the $Q$-criterion. Correspondingly, stretching motions (blue region) are formed at the upstream side of the rotation motions. 
As polymers are added to the flow, they are stretched along the streamlines curved away from the centreline $y=0$ (panel b).
In the Newtonian case at $Re=10$, steady downstream flow separation occurs, forming a pair of weak downstream vortices (identified by the closed streamlines in panel a). As polymers are introduced, the downstream separation region and vortices are suppressed (panel b). Instead, a weak upstream separation zone and the associated upstream vortex pair are formed in response to the increasing elasticity. Similar behaviors have been previously observed in experimental and numerical studies of inertialess viscoelastic flows~\citep{qin2019flow,peng2023numerical}, wherein upstream wake instability is observed, linking to elastic turbulence~\citep{oliveira2001method}. Here we show that this modification also occurs at a finite $Re$.

As $Wi$ increases to $10$ (panel c,f), sheet-like polymeric conformation structures form at a distance of about $0.2$ from the surface of the cylinder. The sheet-like structures resemble those seen in elasto-inertial turbulence (EIT)~\citep{samanta2013elasto,Zhu_XiJNNFM2020, dubief2022} observed in parallel wall-bounded viscoelastic flows at significantly higher $Re$. Interestingly, the location of these polymer sheets is nearly invariant in time and symmetric arrowhead structures can be seen at the channel centre, resembling those of the saturated centre-mode in parallel wall-bounded flows ~\citep{page2020exact}, suggesting a connection between these phenomena.
Fluctuating polymer stress induced by these polymer sheet-like and arrowhead structures can perturb the rotation and stretching motions, resulting in fluctuations in the $Q_v$ field. 
Such a wavy PU state at $(Re,Wi)=(10,10)$ can further evolve into more chaotic stages at $Re=100$, where larger-scale polymer streaks and fragmented rotations develop which break the symmetry of the flow.

%
%
\begin{figure}
    \centering    
    \includegraphics[width=0.7\linewidth, trim=0mm 0mm 0mm 0mm, clip]{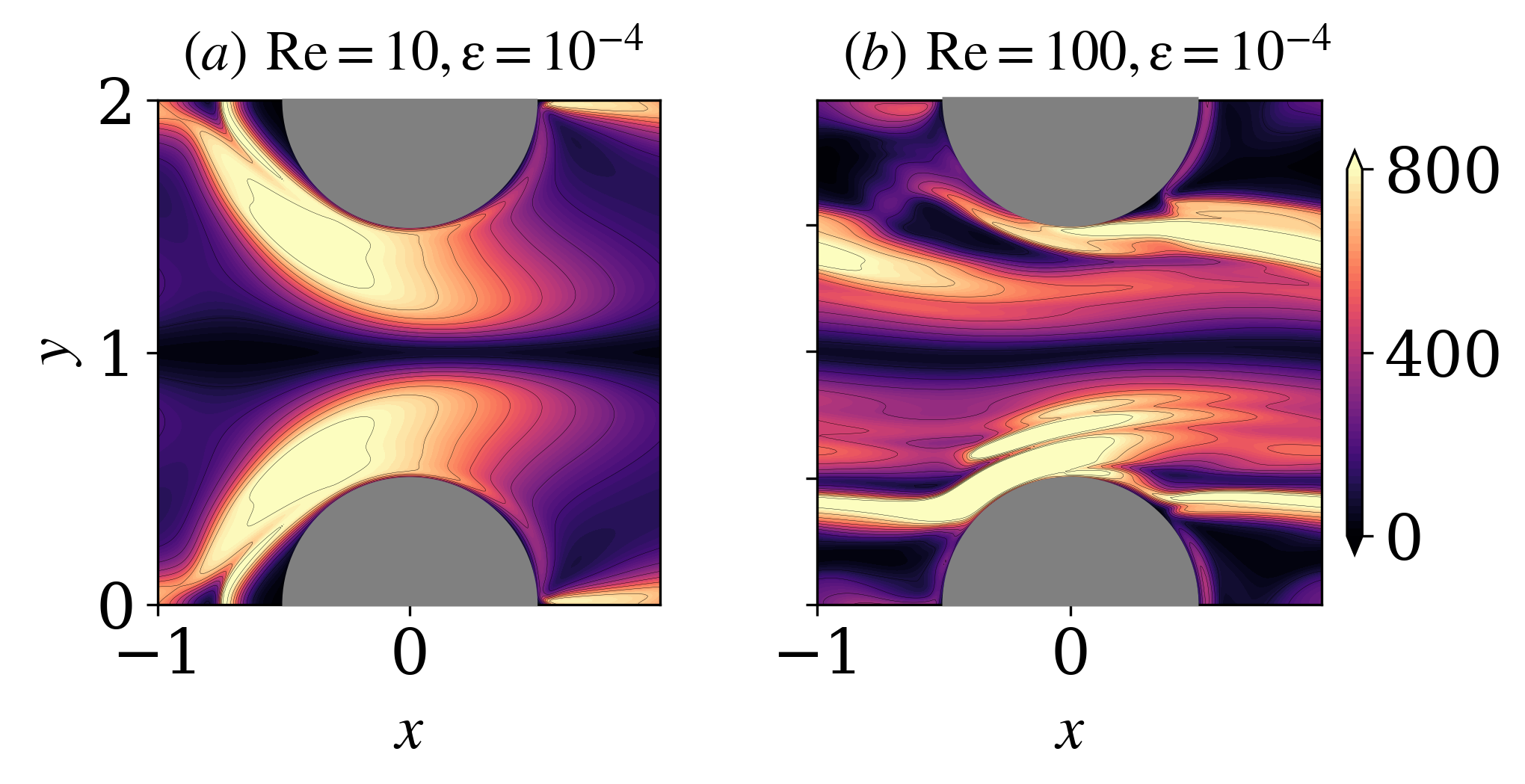}    
    \caption{Snapshots of HD cases ($\mathrm{tr}\,\mbf{\alpha}$ at $\mathrm{\epsilon}=10^{-4}$): (a) $Re=10,Wi=10$,
and (b) $Re=100,Wi=10$.}
    \label{fig:inst_flow_sc}
\end{figure}

\Cref{fig:inst_flow_sc} shows instantaneous snapshots of $\mathrm{tr}\,\mbf{\alpha}$ at the same $Re$ and $Wi$ as \cref{fig:ins_flow}(f,g) but at the larger polymer diffusion of $\epsilon=10^{-4}$. In both cases, we observe the elimination of sharp polymer stress sheets due to the effect of a larger polymer diffusion, despite a similar mean state. As a result, the polymer sheets and arrowhead structures in \cref{fig:ins_flow}(f) are wiped out and a stable state is obtained. As for $Re=100$, despite the influence of the sharp sheets, the unstable flow state is sustained by large-scale streaks which are less sensitive to the polymer diffusion.

%
%
\subsection{Time series}\label{sec:time}

\begin{figure}
      \centering    
      \includegraphics[width=0.7\linewidth, trim=0mm 0mm 0mm 0mm, clip]{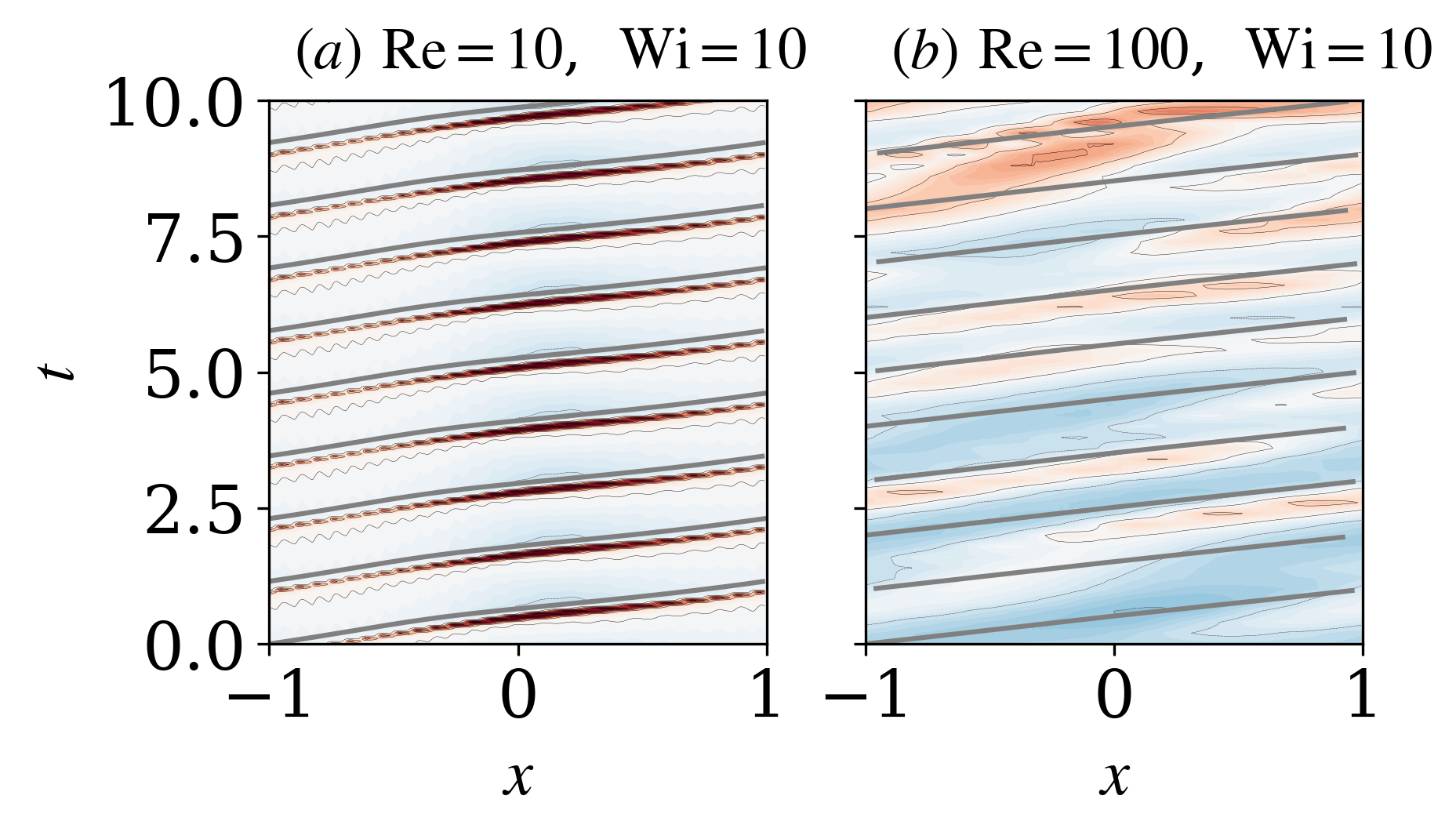}
    \caption{Spatial-temporal $x$-$t$ diagram of normalized polymer fluctuation trace, $\mathrm{tr}\, \mbf{\alpha}^\prime/\sigma $ ($\sigma$ is the standard derivation of $\mathrm{tr}\,\mbf{\alpha}^\prime$), at $y=0.9$: (a) STD, $\mathrm{Re}=10$, $\mathrm{Wi}=10$ and (b) STD, $\mathrm{Re}=100$, $\mathrm{Wi}=10$. The black lines are the particle trajectories of the mean flow, estimating the convective velocity of waves. The discrepancy between the mean flow trajectories and the trajectory of the preparation waves (represented by the contours) indicates the phase speed. }
    \label{fig:xt}
\end{figure}

To investigate the propagation of the wavy PU state in \cref{fig:xt}, we present the spatial-temporal $x$-$t$ diagram of the polymer fluctuation trace, $\mathrm{tr}\,\mbf{\alpha}^\prime$ ($\mbf{\alpha}^\prime=\mbf{\alpha}-\langle \mbf{\alpha} \rangle$), at $y=0.9$, normalized by its standard deviation.
In \cref{fig:xt}, the PU wavy (panel a) and chaotic (panel b) structures propagate downstream. The propagation speed $u_t$ is positive. The phase speed $u_p$ of the wavy structures can be estimated by comparing the convective speed of the mean flow $u_m$ and the propagation speed $u_t$. Therefore, we compute and plot the particle trajectories (grey solid lines) convected by the mean flows in \cref{fig:xt} with their speed approximated as $u_m$. As presented, the wavy trajectory and the particle convective trajectory are closely parallel, suggesting that $u_t$ is approximately equal to $u_m$. Therefore, the $u_p$ of the PU waves is near zero. 

The wavelength $\lambda$ and period $T$ of the wavy structures in the wavy case (panel a) can also be estimated as $\lambda\approx 2$ and $T\approx 1$, respectively, and so are strongly correlated with the periodicity of the flow geometry. 
It can be expected that in the context of single-cylinder flows, where the periodicity of the geometry approaches $\infty$, $\lambda\rightarrow \infty$ and $T\rightarrow \infty$, potentially leading to the absence of such waves. This observation might explain why early transition is not observed in single-cylinder flows. Clearly further investigation of different geometries is needed to verify this. 

%
%

%
%
\section{Nature of instabilities}\label{sec:nature}

So far we have shown the similarity of the instabilities to the centre-mode instability and elasto-inertial turbulence observed in parallel wall-bounded flows~\citep{samanta2013elasto,garg2018viscoelastic,page2020exact}. The centre-mode has been shown to be purely elastic in nature~\citep{khalid2021continuous, Buza2022a,kp2024}.
However, this seems inconsistent with the observation in this study that the PU state depends on  $Re$ (see \cref{fig:regime} for example). In this section, we will further investigate the mechanism of the early unstable PU state.

%
%
\subsection{Budget of turbulent kinetic energy}\label{sec:budget}

To understand the mechanism behind this newly found unstable state in periodic cylinder arrays, we conduct an energy budget analysis of turbulent kinetic energy (TKE) to show its sources and sinks. The time-averaged TKE budget is
\begin{equation}
    \label{eq:budget}
    \frac{ \partial \langle E^\prime\rangle}{\partial t}+\langle \boldsymbol{u} \rangle\cdot\mathbf{\nabla}\langle E^\prime\rangle + \mathbf{\nabla} \cdot \mathcal{T} = \mathcal{P}-\mathcal{D}-\mathcal{W}
\end{equation}
following \citet{Pope_turbulent2000,Zhu_Xi_JNNFM2019}. 
Here, 
\begin{eqnarray}
    \label{eq:prod}
    \mathcal{P}\equiv -\boldsymbol{\nabla}\langle \boldsymbol{u} \rangle : \langle \boldsymbol{u}^\prime \boldsymbol{u}^\prime \rangle,
    \\
    \label{eq:diss}
    \mathcal{D}\equiv \frac{2\beta}{Re}\langle \boldsymbol{S}^\prime : \boldsymbol{S}^\prime \rangle,
    \\
    \label{eq:poly}
    \mathcal{W} \equiv \frac{1-\beta}{Re Wi}\langle \boldsymbol{\tau}_p^\prime : \boldsymbol{S}^\prime \rangle,
\end{eqnarray}
are the shear production, viscous dissipation, and polymer stress contribution of TKE, respectively. $\boldsymbol{S}^\prime\equiv(\partial u^\prime_j/\partial x_i + \partial u^\prime_i/\partial x_j)/2$ is the fluctuating components of the strain-rate tensor.
The last two terms on the left-hand side of \eqref{eq:budget} are convection and energy flux terms that do not contribute to energy exchange and therefore will not be further discussed. 

\begin{figure}
      \centering    
      \includegraphics[width=0.8\linewidth, trim=0mm 0mm 0mm 0mm, clip]{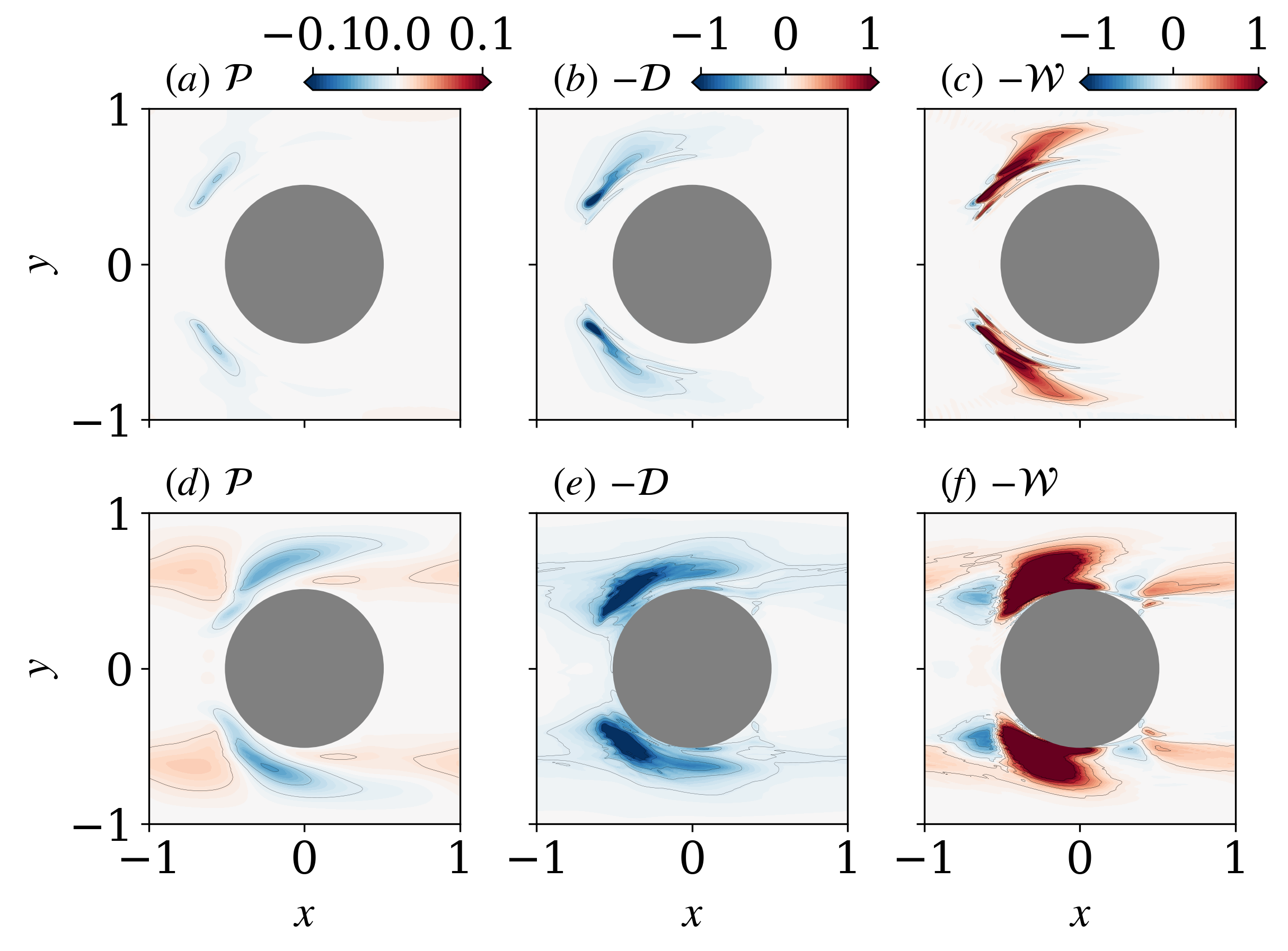}
    \caption{Energy budget of STD wavy case (a-c) at $\mathrm{Re}=10, \mathrm{Wi}=10$ and chaotic case (d-f) at $\mathrm{Re}=100, \mathrm{Wi}=10$.}
    \label{fig:energy}
\end{figure}

\Cref{fig:energy} shows the distribution of $\mathcal{P}$, $-\mathcal{D}$, and $-\mathcal{W}$ in the wavy ($Re=10, Wi=10$) and chaotic ($Re=100, Wi=10$) STD cases. A positive value of these quantities indicates a gain of TKE, and vice versa. Of particular interest, in the wavy case, is the observation of a predominantly positive $-\mathcal{W}$, suggesting the significant contribution of polymer work to the gain of TKE. Therefore, the elastic force mainly drives the instabilities. Meanwhile, the production term 
$\mathcal{P}$ is more than one order of magnitude smaller than the $-\mathcal{W}$ and is, interestingly, negative. This indicates the role of the production term (and, correspondingly, the inertial force) in withdrawing turbulent kinetic energy, exposing $-\mathcal{W}$ as the only source of TKE at a low $Re$. %
As for the chaotic case, despite the overall negative $\mathcal{P}$ and positive $-\mathcal{W}$, weak positive $\mathcal{P}$ and weak negative $-\mathcal{W}$ are observed in localized regions of the flow. This suggests that, although not essential, inertial force can increasingly contribute to the generation of TKE as the flow nears the onset of Newtonian turbulent transition ($Re \approx 150-200$).
The TKE budget analysis strongly supports that the instability observed during the early transition of periodic cylinder arrays is purely/predominantly elastic. This is consistent with the literature~\citep{beneitez2024multistability} that the centre-mode instabilities is elastic. 

In terms of spatial distribution, active TKE gain/sink processes occur in regions with fluctuations of rotation/stretching motions. In the wavy case, significant activities are observed near the edge between the strong rotational and stretching motions (as demonstrated in \cref{fig:ins_flow}) away from the cylinder surface. In the chaotic state, it is closer to the cylinder surface.

%
%
\subsection{Dependence of base state on $Re$}\label{sec:base}

%
%
\begin{figure}
    \centering    
    \includegraphics[width=0.98\linewidth, trim=0mm 0mm 0mm 0mm, clip]{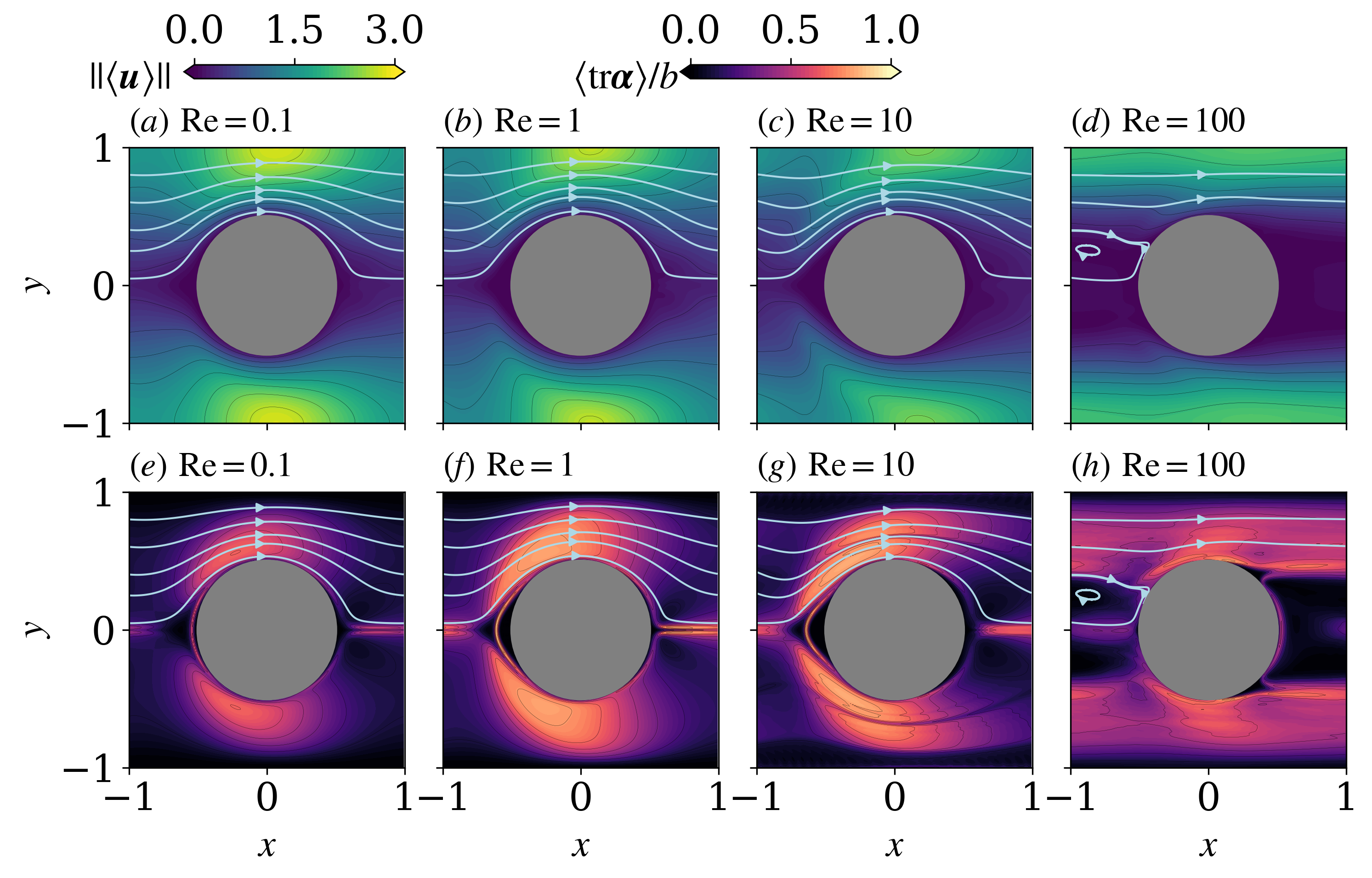}
    \caption{Impact of $Re$ on the base flow states: (a-d) mean velocity magnitude $||\langle\boldsymbol{u}\rangle||$, (e-h) normalised polymer conformation $\langle\mathrm{tr}\mbf{\alpha}\rangle/b$. The blue solid lines represent the instantaneous streamlines of the snapshot.}
    \label{fig:base_re}
\end{figure}

%
%
\begin{figure}
    \centering    
    \includegraphics[width=0.9\linewidth, trim=0mm 0mm 0mm 0mm, clip]{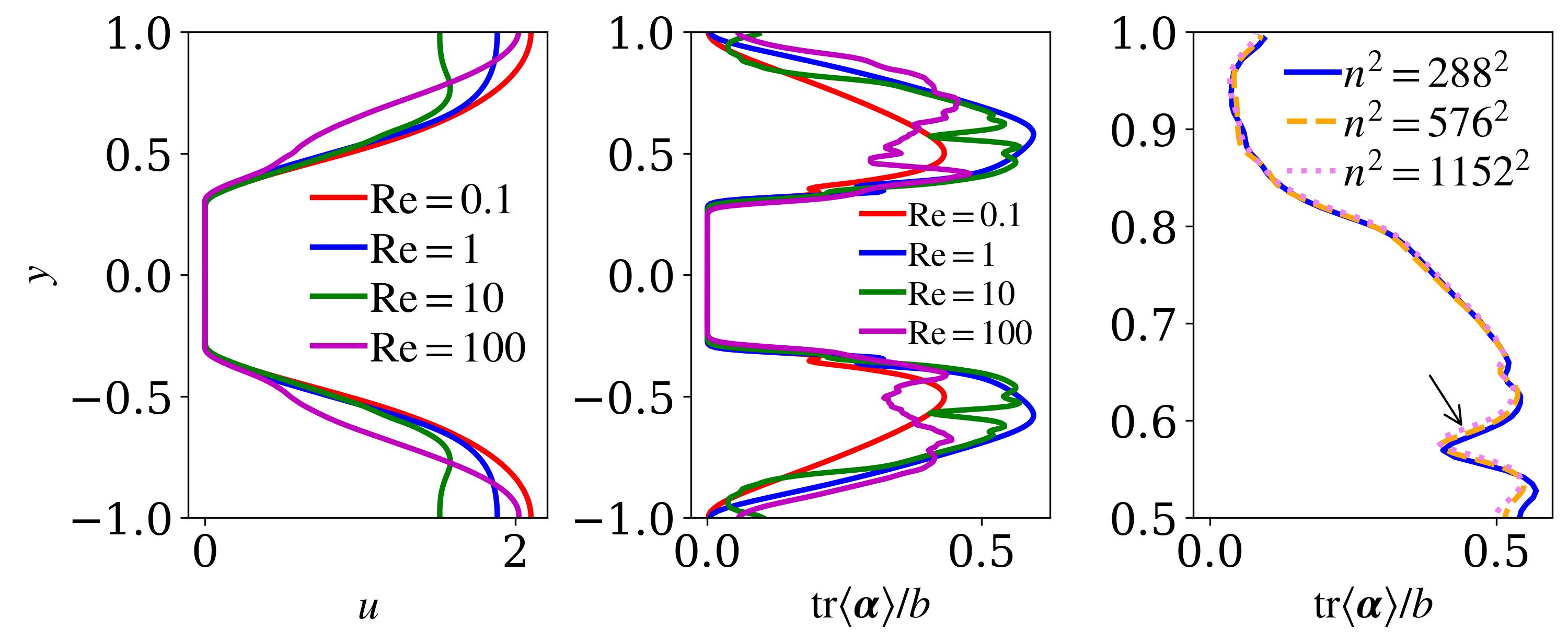}
    \caption{Time-averaged velocity profiles at $x=-0.4$: (a) $\langle u \rangle$, (b) $\langle\mathrm{tr} \mbf{\alpha} \rangle/b$ of cases in~\cref{fig:base_re}, and (c) $\mathrm{tr}\langle \mbf{\alpha} \rangle/b$ of the $(Re,Wi)=(10,10)$ case with different resolution. The arrow points to the invariant polymer sheet.}
    \label{fig:base_profile}
\end{figure}

The TKE budget in \S\ref{sec:budget} suggests the elastic nature of the instabilities in the early turbulence stage of periodic cylinder arrays. Yet it does not explain the $Re$-dependence observed in \cref{fig:regime}. Noticeably, as mentioned by \S\ref{sec:regime}, both elastic and inertial effects can modify the base flow, which can potentially affect the flow stability.

In \cref{fig:base_re}, we show the mean velocity magnitude $ ||\langle\boldsymbol{u}\rangle||$ (`$||\cdot||$' indicates the Euclidean norm) and $\langle \mathrm{tr}\,\mbf{\alpha} \rangle/b$ fields, accompanied by the streamlines, at various $Re$. 
From $Re=0.1$ to $Re=10$, the mean fields are continuously stretched in the streamwise direction. 
Of particular interest, in the unstable $Re=10$ case, a distinct mean polymer sheet is observed on both sides of the cylinder corresponding to the invariant polymer sheets (IPS) observed in \cref{fig:ins_flow}. As will be discussed shortly in \S\ref{sec:pert}, such mean polymer sheets are largely associated with the instabilities of the flow. It is also important to note that these critical polymer sheets are aligned with the mean streamlines, indicating their relation to the mean flow which can be affected by both inertial ($Re$) and elastic ($Wi$) forces. 

The critical polymer sheet divides the flow into two regions with distinct characteristics. Between the critical polymer sheet and the cylinder, the streamlines on the contraction side of the cylinder remain largely parallel, with the polymers fully stretched by the strong near-wall shear. 
This configuration inhibits cross-layer information propagation while facilitating propagation along the layer. Outside the critical polymer sheet, the streamlines exhibit less alignment, thereby permitting more free propagation of information. 

For the $Re=100$ case, the mean flow state differs significantly from the others. The higher $Re$ induces a large downstream separation zone, while the moderate $Wi$ creates an upstream separation zone. These zones connect, forming a dead zone between successive cylinders. Unlike the wavy state, an IPS is not observed, as the chaotic state disrupts most invariant structures.

\Cref{fig:base_profile} shows the time-averaged streamwise velocity $\langle u\rangle$ and normalised polymer conformation trace $\langle\mathrm{tr}\,\mbf{\alpha}\rangle/b$ of cases in \cref{fig:base_re} across the $x=-0.4$ line, representing the active region for TKE generation. While the mean velocity profiles are smooth for all cases, the $\langle\mathrm{tr}\,\mbf{\alpha}\rangle/b$ profiles of the unstable wavy ($Re=10$) and chaotic ($Re=100$) cases exhibit wiggles representing polymer sheets in varying scales. As demonstrated previously, the IPS with a sudden drop in the magnitude of $\langle\mathrm{tr}\,\mbf{\alpha}\rangle/b$, appears at $y=\pm (0.5-0.6)$ in the wavy case. A sensitivity analysis of the grids was conducted, and the independence of these sheets is shown in \cref{fig:base_profile}
(c), suggesting that this observation is not an artifact of insufficient grid resolution. 
In the $Re=100$ case, the polymer sheet expands and forms a wide valley between two peaks in the profile at $y=\pm 0.4$ and $\pm 0.7$.

%
%
\subsection{Evolution of disturbances}\label{sec:pert}

%
%
\begin{figure}
    \centering    
    \includegraphics[width=0.9\linewidth, trim=0mm 0mm 0mm 0mm, clip]{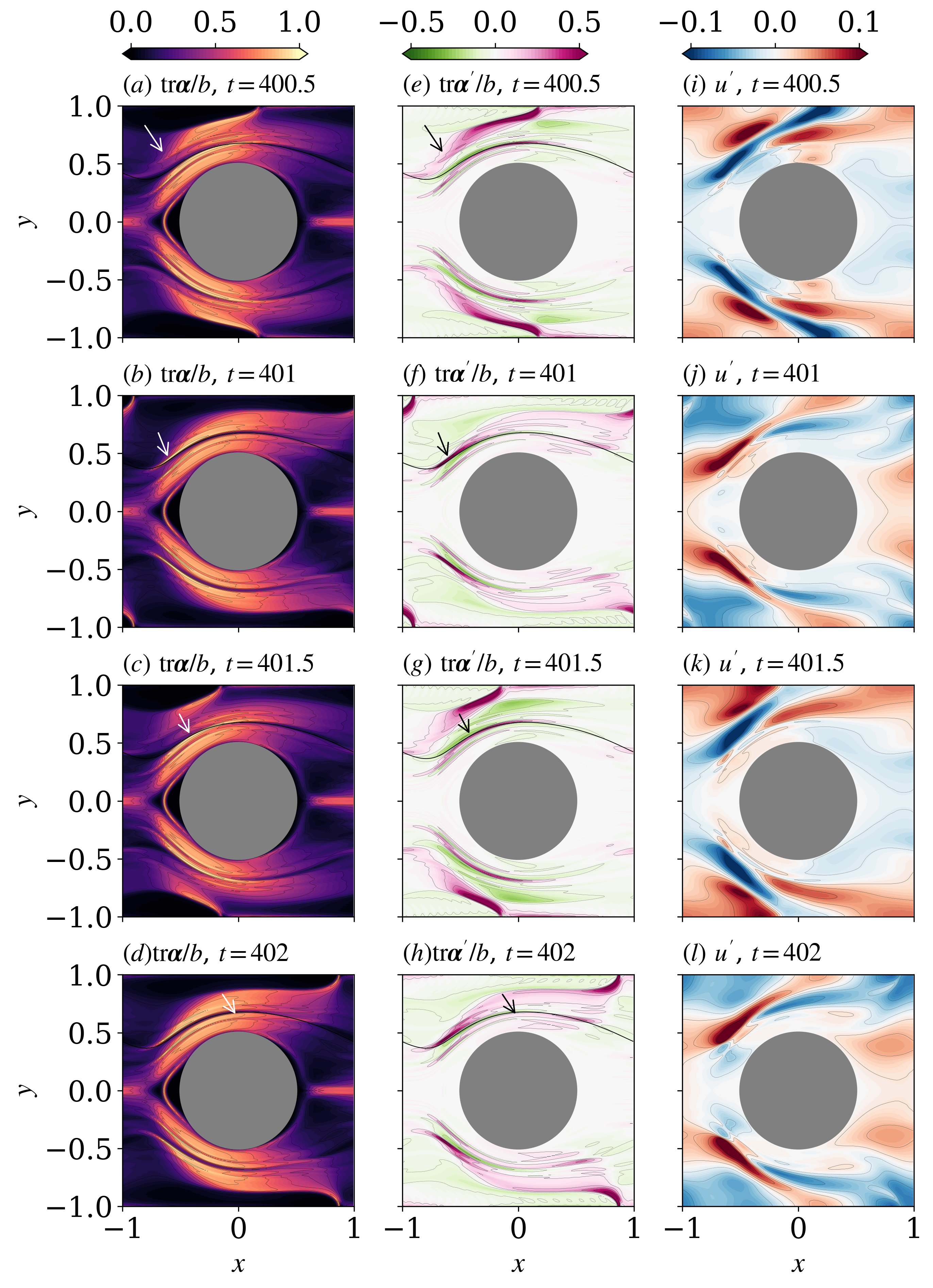}
    \caption{Evolution of flow disturbances: instantaneous snapshots of normalised polymer extension fluctuation $\mathrm{tr}\mbf{\alpha}^\prime/b$ (first row) and streamwise velocity fluctuation $u^\prime$ (second row) at $(Re,Wi)=(10,10)$ for STD cases. The arrow indicates the position of a progressively developing disturbance in polymer conformation. The grey solid line represents the mean invariant polymer sheet observed in  \cref{fig:base_profile} (c).} %
    \label{fig:FFperb}
\end{figure}

The IPS plays a critical role in generating and maintaining the instabilities of the flow. Phenomenologically, it acts as an unstable manifold in that polymer structures are stretched along the trajectory aligned with the polymer sheets. This can be seen in \cref{fig:FFperb}, which displays the transient snapshots of normalised polymer conformation trace $\mathrm{tr}\,\mbf{\alpha}/b$ and the disturbance in $\mathrm{tr}\,\mbf{\alpha}/b$ and the streamwise velocity $u$.  
From $t=400.5$ to $t=402$, the IPS (indicated by the solid line in \cref{fig:FFperb}(e-h)) maintains its shape and remains relatively invariant over time.
The elastic wave, indicated by the intense purple region in \cref{fig:FFperb}(e-h), propagates downstream in the region outside the invariant sheet. 
As the wavy structure passes the cylinder, the tail of the wavy structure (indicated by the arrow in panel a,e) encounters the invariant sheet, resulting in significant vertical compression (panel f) and elongation along the streamlines (panel g). This occurs because the parallel streamlines between the invariant sheet and the cylinder inhibit cross-layer information propagation.
While the polymer is stretching along the streamlined direction, the subsequent wavefront pushes relaxed polymers (colored green) toward the invariant sheet (panel g). 
These polymers then continue to relax along the streamlines (panel h).
The stretching/recoiling process at the IPS induces significant velocity fluctuations, leading to intensive TKE production at the point where the wave and sheet intersect. This picture aligns with the TKE budget in \cref{fig:energy}, indicating that elastic forces are primarily responsible for TKE generation.
Note also that, due to the effect of the invariant sheet, the flow inside the polymer sheets exhibits a relatively calm state. Perturbations are largely suppressed in this region.

%
%
\section{Polymeric concentration and maximal polymer extension }

In previous sections, we mainly explore the influence of $Re$, $Wi$, and $\epsilon$ (or $Sc$) on the early turbulence and the drag enhancement of the flow. The rest of the parameters, i.e., the polymer maximal extension parameter $b$ and the polymer concentration $\beta$ are fixed to $b=1000$ and $\beta=0.9$. To investigate the potential impact of $b$ and $\beta$, we conduct two additional `LE' and `VC' cases (listed in \cref{tab:simul_overview}) at $Re=10$ in which $b$ and $\beta$ are set to 6400 and 0.7 or 0.97, respectively. 

\begin{figure}
    \centering    
    \includegraphics[width=0.98\linewidth, trim=0mm 0mm 0mm 0mm, clip]{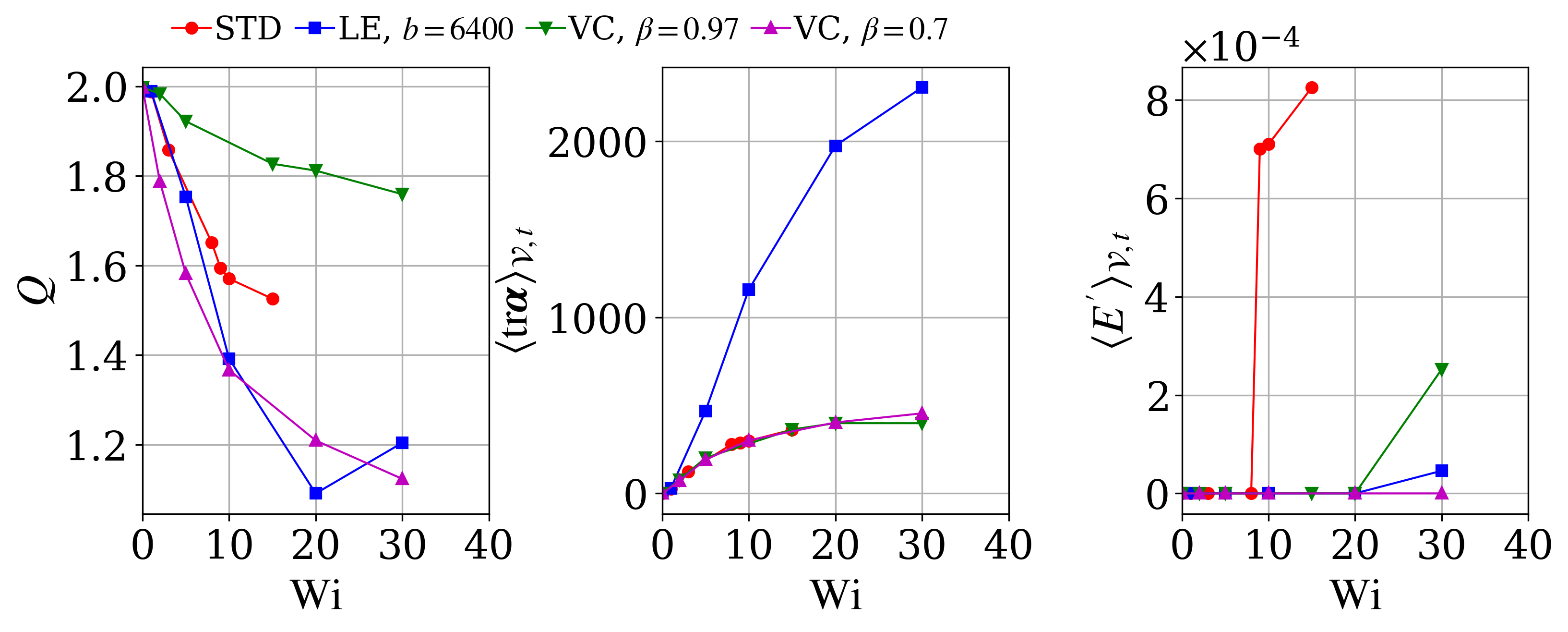}
    \caption{The impact on flow statistics due to different $b$ and $\beta$ used in cases LE and VC: (a) volumetric flux, (b) trace of polymer conformation tensor, (c) turbulent kinetic energy.}
    \label{fig:stat_bbeta}
\end{figure}

\begin{figure}
    \centering    
    \includegraphics[width=0.95\linewidth, trim=0mm 0mm 0mm 0mm, clip]{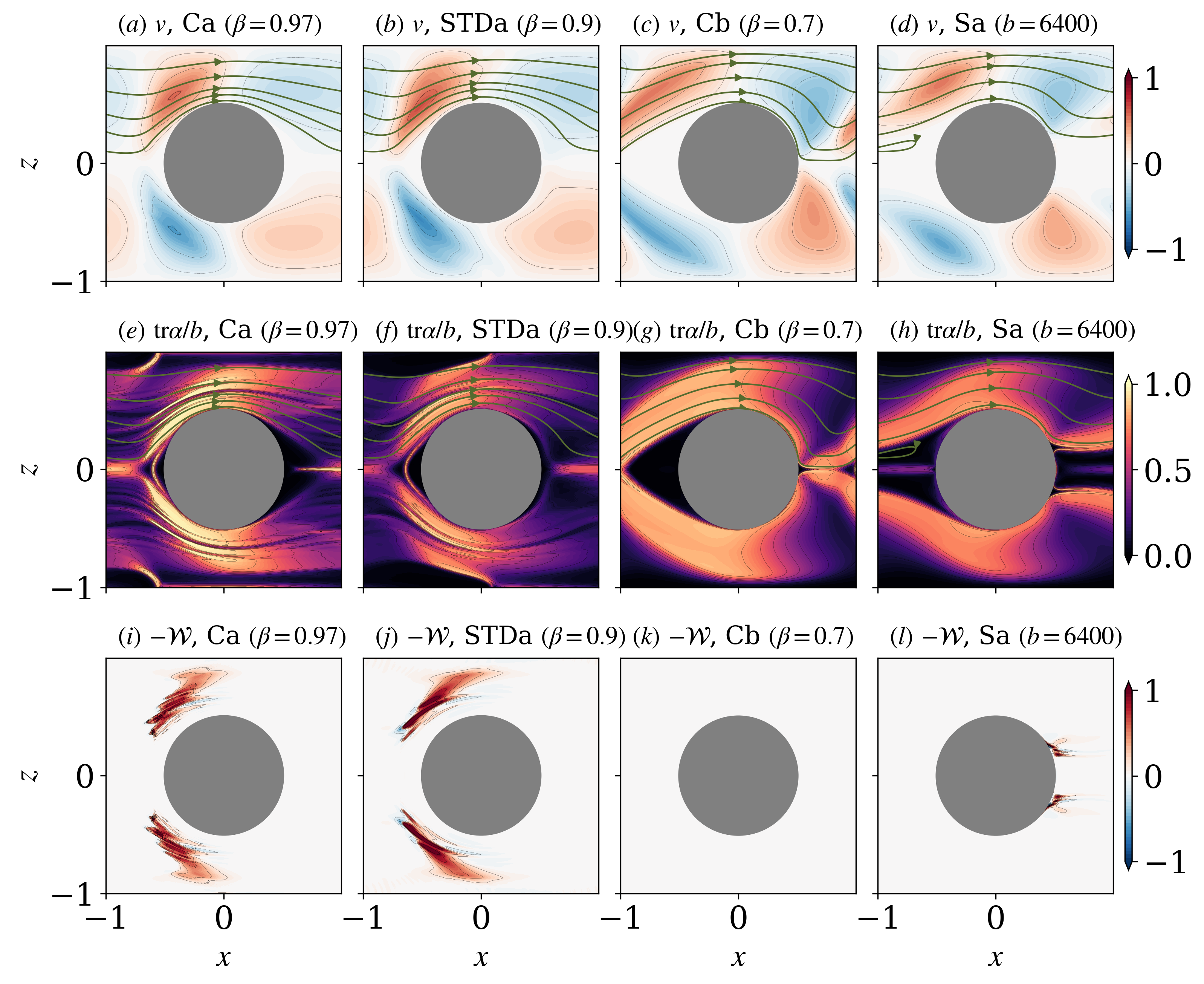}
    \caption{Impact of instantaneous flow field by $b$ and $\beta$: (a-d) $y$-component of velocity, (e-h) normalised polymer extension $\mathrm{tr}\,\mbf{\alpha}/b$ and (i-l) polymer work of TKE budget. }

    \label{fig:inst_flow_bbeta}
\end{figure}

\Cref{fig:stat_bbeta} shows the statistics of these cases alongside the corresponding STD at the same $Re$. The onset of PL stage is independent of $b$ and $\beta$ and remains at $Wi$ $\approx 2$. 
However, the onset of PU depends sensitively on $b$ and $\beta$ due to the capability of elastic forces to modify the base state. Intuitively, we expect that the increasing $b$ and decreasing $\beta$ will amply the polymer force and promote the chaotic PU state. 
This is consistent with case VC at $\beta=0.97$ in which the onset of PU ($\mathrm{Wi}\approx 20$) is delayed compared with the STD case ($\beta=0.9$). 
However, it is inconsistent with the observations in case LE at $b=6400$ and case VC at $\beta=0.7$. In VC, a delay in the PU state (compared with the STD case) is observed up to $Wi=30$, whereas in LE, the PU state is absent at the highest $Wi=30$ studied. 
Also, it is interesting to note that in the LE cases, the volumetric flux profile rises from $Wi=20$ to $30$. This phenomenon is attributed to the formation of a dead zone between cylinders under large elastic forces, as will be presented in \cref{fig:inst_flow_bbeta}. 

%
%

To clearly demonstrate the modification of the flow state by $b$ and $\beta$, we show the instantaneous spanwise velocity component $v$, normalised polymer extension $\mathrm{tr}\, \mbf{\alpha}/b$, and the polymer work in the TKE budget in \cref{fig:inst_flow_bbeta}. Four cases are chosen and named as (i) VCa: $(Re,Wi,\boldsymbol{\beta},b)=(10,30,\boldsymbol{0.97},1000)$ (ii) STDa: $(Re,Wi,\boldsymbol{\beta},b)=(10,10,\boldsymbol{0.9},1000)$, (iii) VCb: $(Re,Wi,\boldsymbol{\beta},b)=(10,30,\boldsymbol{0.7},1000)$, and (iv) LEa: $(Re,Wi,\beta,\boldsymbol{b})=(10,30,0.9,\boldsymbol{6400})$.

As $\beta$ decreases from $0.97$ (VCa) to $0.9$ (STDa), the base state exhibits considerable similarity, particularly in terms of finite upstream separation regions near $y=0$ adjacent to the cylinder. The flow fields are subtly modified, with the upstream vortex expanding and downstream vortex being suppressed as $\beta$ increases.
Importantly, both polymer sheets and arrowhead structures are observed in these cases, indicating a consistent transition mechanism.

As $\beta$ decreases to $0.7$ (VCb) from $0.9$ (STDa) or $b$ increases to $6400$ (LEa), the base state undergoes significant modification. The flow structures are stretched along the $x$-axis, creating an extensive upstream separation zone. 
Furthermore, in the LEa case, the upstream separation zone connects the upstream cylinder, creating a dead zone. 
The intensive elastic force markedly reshapes the flow near the upstream wake, leading to the formation of a smooth and thick polymer-modified boundary layer, characterized by parallel, less curved streamlines and fully stretched polymers. This modification potentially discourages instabilities. This region can be seen as an expansion of the area between the invariant sheet and cylinder observed in VCa and STDa cases. However, due to the polymer-induced reshaping of the mean flow, neither the IPS nor the unstable arrowhead waves appear, resulting in a smooth polymer field across the channel. Therefore, a stable flow state is achieved at the VCb cases, despite the elastic forces being the driver of the instabilities. Interestingly, in the LEa case, the flow remains unstable due to the vibration of birefringent strands (regions of fully extended polymer downstream of the cylinder~\citep{haward2021bifurcations}).

In \cref{fig:inst_flow_bbeta}(i-j), we also compare the polymer work. All the unstable cases (i.e., STDa, VCa, and LEa) have a dominantly positive $-\mathcal{W}$, confirming the elastic nature of the instabilities. However, the active location of polymer work between panels (i,j) and (l) is different. While STDa and Ca cases have an active region at the upstream side of the cylinder, the active regions in the LEa cases appear at the downstream wake. As previously discussed, the TKE in the VCa and STDa cases is attributed to the interaction between the IPS and the upcoming waves. In contrast, in the LEa case, it results from the interaction between birefringent strands and the cylinder.

%
%

%
%
\section{Discussion
}\label{sec:disc}

%
%
In this study, we conducted 2D numerical simulations to investigate the viscoelastic flow past a  periodic cylinder array at finite Reynolds numbers. This reveals a multi-stage transition to an early chaotic wavy state consisting of
\begin{itemize}
    \item A Newtonian-like Laminar state (NL) with flow statistics resembling the Newtonian laminar flow
    \item A Polymer-modified Laminar (PL) state where  the base flow fields are significantly modified by polymers and drag enhancement is seen.
    \item A Polymer-induced unstable (PU) state, with elastic-driven instabilities and early transition to a wavy chaotic state. 
\end{itemize}

%
%
Upon changing the polymer diffusivity $\epsilon$, we found that the existence of the elastic-driven instabilities and wavy chaos relies heavily on a small $\epsilon$. At $\epsilon=10^{-5}$, the PU state is triggered at $Re=10$ (and $Wi=9$) which is significantly lower than the Newtonian flows of $Re=150-200$, leading to an early transition. 
As $\epsilon$ increases to $10^{-4}$, the border of the early instabilities shifts to the upper-right in the $Re$-$Wi$ parameter space, requiring a higher $Re=30$ ($Wi=15$) to induce transition.
In contrast, the onset of PL is insensitive to both $\epsilon$ and $Re$, occurring at $Wi=2$-$3$. This insensitivity suggests the pure elastic nature of the PL state.
%
%
Diagnosing features of the unstable wavy flows (STD, $Re=10, Wi=10$) in the PU state reveals arrowhead and sheet-like polymer stress structures reminiscent of the saturated centre-mode instability and elasto-inertial turbulence found in parallel wall-bounded flows~\citep{page2020exact}. This has not been reported before in this context presumably because of the low values of $\epsilon$ (large values of $Sc$) used here for the first time.
Furthermore, typical ET with downstream/upstream symmetric breaking of wakes~\citep{haward2021bifurcations,qin2019flow} has not been observed. 
This might be because of the restricted geometry studied, which prevents the growth of the upstream and downstream separation.  The fact that the arrowhead and polymer sheet flows cannot be seen much below $Re \approx 10$ suggests that EIT may not be directly connected with ET although this needs to be checked in a more extended domain.

%
%
The instabilities responsible for triggering and sustaining the PU state are fundamentally elastic in nature. This conclusion is supported by the TKE budget analysis, where the polymer work term $-\mathcal{W}$ is predominantly positive,  indicating that elastic forces contribute significantly to the production of turbulent kinetic energy (TKE).
In contrast, the shear production term $\mathcal{P}$ is weak and mostly negative, suggesting that inertial forces play a  minor role in withdrawing TKE. 
We show that the TKE is produced by the interaction between the invariant polymer sheet (IPS) and the upcoming elastic waves. This interaction causes intensive stretching and recoiling of polymers along the polymer sheet, subsequently inducing velocity fluctuations.
The $Re$-dependence of the PU state is likely attributed to the substantial modification of the base flow state by inertial forces, which eradicate the IPS. 

Exploring different polymer maximum extension $b$ and concentration $\beta$ reveals the significant role of elastic forces in modifying the base state, thereby inducing drag enhancement. Notably, due to the varying modifications of the base state as $b$ and $\beta$ are altered, the unstable PU state is not linearly dependent on the elastic force, despite these instabilities being purely elastic. Rather, further increase of the elastic force (achieved by either increasing $b$ or decreasing $\beta$) could suppress the unstable state. A comprehensive investigation of the instability diagram in a broader parameter space across an expanded parameter space of $b$, $\beta$ and $\epsilon$ is essential to fully understand the complex relationship between the base state modifications and the onset of instabilities. We will leave this for the future.

\bibliographystyle{jfm}
\bibliography{References.bib,polymer.bib}

\begin{thebibliography}{70}
\expandafter\ifx\csname natexlab\endcsname\relax\def\natexlab#1{#1}\fi
\def\au#1{#1} \def\ed#1{#1} \def\yr#1{#1}\def\at#1{#1}\def\jt#1{\textit{#1}}
  \def\bt#1{#1}\def\bvol#1{\textbf{#1}} \def\vol#1{#1} \def\pg#1{#1}
  \def\publ#1{#1}\def\arxiv#1{#1}\def\org#1{#1}\def\st#1{\textit{#1}}

\bibitem[Alves {\em et~al.\/}(2021)Alves, Oliveira \&
  Pinho]{alves2021numerical}
{\sc \au{Alves, M.~A.}, \au{Oliveira, P.~J.} \& \au{Pinho, F.~T.}} \yr{2021}
  \at{Numerical methods for viscoelastic fluid flows}  \bvol{53},
  \pg{509--541}.

\bibitem[Angot(1999)]{angot1999analysis}
{\sc \au{Angot, P.}} \yr{1999}  \at{Analysis of singular perturbations on the
  brinkman problem for fictitious domain models of viscous flows}.  \jt{Math.
  meth. in the Appl. Sci.}  \bvol{22}~(16),  \pg{1395--1412}.

\bibitem[Ascher {\em et~al.\/}(1997)Ascher, Ruuth \&
  Spiteri]{ascher1997implicit}
{\sc \au{Ascher, U.~M.}, \au{Ruuth, S.~J.} \& \au{Spiteri, R.~J.}} \yr{1997}
  \at{Implicit-explicit runge-kutta methods for time-dependent partial
  differential equations}.  \jt{Appl. Numer. Math.}  \bvol{25}~(2-3),
  \pg{151--167}.

\bibitem[Azaiez \& Homsy(1994)]{azaiez1994numerical}
{\sc \au{Azaiez, J.} \& \au{Homsy, G.~M.}} \yr{1994}  \at{Numerical simulation
  of non-newtonian free shear flows at high reynolds numbers}.  \jt{J.
  Non-Newton. Fluid Mech.}  \bvol{52}~(3),  \pg{333--374}.

\bibitem[Beneitez {\em et~al.\/}(2024)Beneitez, Page, Dubief \&
  Kerswell]{beneitez2024multistability}
{\sc \au{Beneitez, M.}, \au{Page, J.}, \au{Dubief, Y.} \& \au{Kerswell, R.~R.}}
  \yr{2024}  \at{Multistability of elasto-inertial two-dimensional channel
  flow}.  \jt{J. Fluid Mech.}  \bvol{981},  \pg{A30}.

\bibitem[Beneitez {\em et~al.\/}(2023)Beneitez, Page \& Kerswell]{beneitez2023}
{\sc \au{Beneitez, M.}, \au{Page, J.} \& \au{Kerswell, R.~R.}} \yr{2023}
  \at{Polymer diffusive instability leading to elastic turbulence in plane
  couette flow}.  \jt{Phys. Rev. Fluids}  \bvol{8},  \pg{L101901}.

\bibitem[Beris \& Mavrantzas(1994)]{beris1994compatibility}
{\sc \au{Beris, A.~N.} \& \au{Mavrantzas, V.~G.}} \yr{1994}  \at{On the
  compatibility between various macroscopic formalisms for the concentration
  and flow of dilute polymer solutions}.  \jt{J. Rheol.}  \bvol{38}~(5),
  \pg{1235--1250}.

\bibitem[Bird {\em et~al.\/}(1987)Bird, Hassager, Armstrong \&
  Crurtis]{bird1987dynamics}
{\sc \au{Bird, R.~B.}, \au{Hassager, O.}, \au{Armstrong, R.~C.} \& \au{Crurtis,
  C.~F.}} \yr{1987} {\em Dynamics of polymeric liquids. Vol. 2 Kinetic
  Theory\/}.  \publ{John Wiley and Sons Inc., New York, NY}.

\bibitem[Burns {\em et~al.\/}(2020)Burns, Vasil, Oishi, Lecoanet \&
  Brown]{burns2020dedalus}
{\sc \au{Burns, K.~J.}, \au{Vasil, G.~M.}, \au{Oishi, J.~S.}, \au{Lecoanet, D.}
  \& \au{Brown, B.~P.}} \yr{2020}  \at{Dedalus: A flexible framework for
  numerical simulations with spectral methods}.  \jt{Phys. Rev. Res.}
  \bvol{2}~(2),  \pg{023068}.

\bibitem[Buza {\em et~al.\/}(2022{\natexlab{{\em a\/}}})Buza, Beneitez, Page \&
  Kerswell]{buza2022finite}
{\sc \au{Buza, G.}, \au{Beneitez, M.}, \au{Page, J.} \& \au{Kerswell, R.~R.}}
  \yr{2022{\natexlab{{\em a\/}}}}  \at{Finite-amplitude elastic waves in
  viscoelastic channel flow from large to zero reynolds number}.  \jt{J. Fluid
  Mech.}  \bvol{951},  \pg{A3}.

\bibitem[Buza {\em et~al.\/}(2022{\natexlab{{\em b\/}}})Buza, Page \&
  Kerswell]{Buza2022a}
{\sc \au{Buza, G.}, \au{Page, J.} \& \au{Kerswell, R.~R}}
  \yr{2022{\natexlab{{\em b\/}}}}  \at{Weakly nonlinear analysis of the
  viscoelastic instability in channel flow for finite and vanishing reynolds
  numbers}.  \jt{J. Fluid Mech.}  \bvol{940},  \pg{A11}.

\bibitem[Cadot \& Kumar(2000)]{cadot2000experimental}
{\sc \au{Cadot, O.} \& \au{Kumar, S.}} \yr{2000}  \at{Experimental
  characterization of viscoelastic effects on two-and three-dimensional shear
  instabilities}.  \jt{J. Fluid Mech.}  \bvol{416},  \pg{151--172}.

\bibitem[Choueiri {\em et~al.\/}(2021)Choueiri, Lopez, Varshney, Sankar \&
  Hof]{choueiri2021experimental}
{\sc \au{Choueiri, G.~H.}, \au{Lopez, J.~M.}, \au{Varshney, A.}, \au{Sankar,
  S.} \& \au{Hof, B.}} \yr{2021}  \at{Experimental observation of the origin
  and structure of elastoinertial turbulence}.  \jt{Proc. Natl. Acad. Sci. U.
  S. A.}  \bvol{118}~(45),  \pg{e2102350118}.

\bibitem[Couchman {\em et~al.\/}(2024)Couchman, Beneitez, Page \&
  Kerswell]{Couchman2024}
{\sc \au{Couchman, M. M.~P.}, \au{Beneitez, M.}, \au{Page, J.} \& \au{Kerswell,
  R.~R.}} \yr{2024}  \at{Inertial enhancement of the polymer diffusive
  instability}.  \jt{J. Fluid Mech.}  \bvol{981},  \pg{A2}.

\bibitem[Datta {\em et~al.\/}(2022)Datta, Ardekani, Arratia, Beris,
  Bischofberger, McKinley, Eggers, L{\'o}pez-Aguilar, Fielding, Frishman {\em
  et~al.\/}]{datta2022}
{\sc \au{Datta, S.~S.}, \au{Ardekani, A.~M.}, \au{Arratia, P.~E.}, \au{Beris,
  A.~N.}, \au{Bischofberger, I.}, \au{McKinley, G.~H.}, \au{Eggers, J.~G.},
  \au{L{\'o}pez-Aguilar, J.~E.}, \au{Fielding, S.~M.}, \au{Frishman, A.} \&
  \au{others}} \yr{2022}  \at{Perspectives on viscoelastic flow instabilities
  and elastic turbulence}.  \jt{Phys. Rev. Fluids}  \bvol{7}~(8),  \pg{080701}.

\bibitem[Dubief {\em et~al.\/}(2022)Dubief, Page, Kerswell, Terrapon \&
  Steinberg]{dubief2022}
{\sc \au{Dubief, Y.}, \au{Page, J.}, \au{Kerswell, R.~R.}, \au{Terrapon, V.~E.}
  \& \au{Steinberg, V.}} \yr{2022}  \at{First coherent structure in
  elasto-inertial turbulence}.  \jt{Phys. Rev. Fluids}  \bvol{7}~(7),
  \pg{073301}.

\bibitem[Dubief {\em et~al.\/}(2013)Dubief, Terrapon \&
  Soria]{dubief2013mechanism}
{\sc \au{Dubief, Y.}, \au{Terrapon, V.E.} \& \au{Soria, J.}} \yr{2013}  \at{On
  the mechanism of elasto-inertial turbulence}.  \jt{Phys. Fluids}
  \bvol{25}~(11).

\bibitem[Dubief {\em et~al.\/}(2023)Dubief, Terrapon \& Hof]{dubief2023elasto}
{\sc \au{Dubief, Y.}, \au{Terrapon, V.~E.} \& \au{Hof, B.}} \yr{2023}
  \at{Elasto-inertial turbulence}.  \jt{Ann. Rev. Fluid Mech.}  \bvol{55},
  \pg{675--705}.

\bibitem[Fisher {\em et~al.\/}(1996)Fisher, Torrance \&
  Sikka]{fisher1996analysis}
{\sc \au{Fisher, T.S.}, \au{Torrance, K.E.} \& \au{Sikka, K.K.}} \yr{1996}
  Analysis and optimization of a natural draft heat sink system.  \bt{In {\em
  InterSociety Conference on Thermal Phenomena in Electronic Systems, I-THERM
  V\/}},  \pg{pp. 115--123}.

\bibitem[Gadd(1966)]{gadd1966effects}
{\sc \au{Gadd, G.~E.}} \yr{1966}  \at{Effects of long-chain molecule additives
  in water on vortex streets}.  \jt{Nature}  \bvol{211}~(5045),  \pg{169--170}.

\bibitem[Garg {\em et~al.\/}(2018)Garg, Chaudhary, Khalid, Shankar \&
  Subramanian]{garg2018viscoelastic}
{\sc \au{Garg, P.}, \au{Chaudhary, I.}, \au{Khalid, M.}, \au{Shankar, V.} \&
  \au{Subramanian, G.}} \yr{2018}  \at{Viscoelastic pipe flow is linearly
  unstable}.  \jt{Phys. Rev. Lett.}  \bvol{121}~(2),  \pg{024502}.

\bibitem[Grilli {\em et~al.\/}(2013)Grilli, V{\'a}zquez-Quesada \&
  Ellero]{grilli2013transition}
{\sc \au{Grilli, M.}, \au{V{\'a}zquez-Quesada, A.} \& \au{Ellero, M.}}
  \yr{2013}  \at{Transition to turbulence and mixing in a viscoelastic fluid
  flowing inside a channel with a periodic array of cylindrical obstacles}.
  \jt{Phys. Rev. Lett.}  \bvol{110}~(17),  \pg{174501}.

\bibitem[Groisman \& Steinberg(2000)]{groisman2000elastic}
{\sc \au{Groisman, A.} \& \au{Steinberg, V.}} \yr{2000}  \at{Elastic turbulence
  in a polymer solution flow}.  \jt{Nature}  \bvol{405}~(6782),  \pg{53--55}.

\bibitem[Haward {\em et~al.\/}(2021)Haward, Hopkins, Varchanis \&
  Shen]{haward2021bifurcations}
{\sc \au{Haward, S.~J.}, \au{Hopkins, C.~C.}, \au{Varchanis, S.} \& \au{Shen,
  A.~Q.}} \yr{2021}  \at{Bifurcations in flows of complex fluids around
  microfluidic cylinders}.  \jt{Lab on a Chip}  \bvol{21}~(21),
  \pg{4041--4059}.

\bibitem[Hester {\em et~al.\/}(2021)Hester, Vasil \&
  Burns]{hester2021improving}
{\sc \au{Hester, E.~W.}, \au{Vasil, G.~M.} \& \au{Burns, K.~J.}} \yr{2021}
  \at{Improving accuracy of volume penalised fluid-solid interactions}.  \jt{J.
  Chem. Phys.}  \bvol{430},  \pg{110043}.

\bibitem[Hopkins {\em et~al.\/}(2020)Hopkins, Haward \&
  Shen]{hopkins2020purely}
{\sc \au{Hopkins, C.~C.}, \au{Haward, S.~J.} \& \au{Shen, A.~Q.}} \yr{2020}
  \at{Purely elastic fluid--structure interactions in microfluidics:
  implications for mucociliary flows}.  \jt{Small}  \bvol{16}~(9),
  \pg{1903872}.

\bibitem[Housiadas \& Beris(2003)]{housiadas2003polymer}
{\sc \au{Housiadas, K.~D.} \& \au{Beris, A.~N.}} \yr{2003}  \at{Polymer-induced
  drag reduction: Effects of the variations in elasticity and inertia in
  turbulent viscoelastic channel flow}.  \jt{Physics of Fluids}  \bvol{15}~(8),
   \pg{2369--2384}.

\bibitem[James \& Gupta(1971)]{james1971drag}
{\sc \au{James, D.~F.} \& \au{Gupta, O.~P.}} \yr{1971} Drag on circular
  cylinders in dilute polymer solutions.  \bt{In {\em Chem. Eng. Prog. Symp.
  Ser\/}}, ,  \vol{vol.~67},  \pg{pp. 62--73}.

\bibitem[Kenney {\em et~al.\/}(2013)Kenney, Poper, Chapagain \&
  Christopher]{kenney2013large}
{\sc \au{Kenney, S.}, \au{Poper, K.}, \au{Chapagain, G.} \& \au{Christopher,
  G.~F.}} \yr{2013}  \at{Large deborah number flows around confined
  microfluidic cylinders}.  \jt{Rheol. Acta}  \bvol{52},  \pg{485--497}.

\bibitem[Kerswell \& Page(2024)]{kp2024}
{\sc \au{Kerswell, R.~R.} \& \au{Page, J.}} \yr{2024}  \at{Asymptotics of the
  centre-mode instability in viscoelastic channel flow: with and without
  inertia}.  \jt{J. Fluid Mech.}  \bvol{991},  \pg{A13}.

\bibitem[Khalid {\em et~al.\/}(2021{\natexlab{{\em a\/}}})Khalid, Chaudhary,
  Garg, Shankar \& Subramanian]{Khalid2021a}
{\sc \au{Khalid, M.}, \au{Chaudhary, I.}, \au{Garg, P.}, \au{Shankar, V.} \&
  \au{Subramanian, G.}} \yr{2021{\natexlab{{\em a\/}}}}  \at{The centre-mode
  instability of viscoelastic plane {P}oiseuille flow}.  \jt{J. Fluid Mech.}
  \bvol{915},  \pg{A43}.

\bibitem[Khalid {\em et~al.\/}(2021{\natexlab{{\em b\/}}})Khalid, Shankar \&
  Subramanian]{khalid2021continuous}
{\sc \au{Khalid, M.}, \au{Shankar, V.} \& \au{Subramanian, G.}}
  \yr{2021{\natexlab{{\em b\/}}}}  \at{Continuous pathway between the
  elasto-inertial and elastic turbulent states in viscoelastic channel flow}.
  \jt{Phys. Rev. Lett.}  \bvol{127}~(13),  \pg{134502}.

\bibitem[Khomami \& Moreno(1997)]{khomami1997stability}
{\sc \au{Khomami, B.} \& \au{Moreno, L.~D.}} \yr{1997}  \at{Stability of
  viscoelastic flow around periodic arrays of cylinders}.  \jt{Rheologica acta}
   \bvol{36},  \pg{367--383}.

\bibitem[Kim {\em et~al.\/}(2008)Kim, Adrian, Balachandar \&
  Sureshkumar]{kim2008dynamics}
{\sc \au{Kim, K.}, \au{Adrian, R.~J.}, \au{Balachandar, S.} \& \au{Sureshkumar,
  R.}} \yr{2008}  \at{Dynamics of hairpin vortices and polymer-induced
  turbulent drag reduction}.  \jt{Phys. Rev. Lett.}  \bvol{100}~(13),
  \pg{134504}.

\bibitem[Kim {\em et~al.\/}(2007)Kim, Li, Sureshkumar, Balachandar \&
  Adrian]{kim2007effects}
{\sc \au{Kim, K.}, \au{Li, C.~F.}, \au{Sureshkumar, R.}, \au{Balachandar, S.}
  \& \au{Adrian, R.~J.}} \yr{2007}  \at{Effects of polymer stresses on eddy
  structures in drag-reduced turbulent channel flow}.  \jt{J. Fluid Mech.}
  \bvol{584},  \pg{281--299}.

\bibitem[Layec \& Layec-Raphalen(1983)]{layec1983instability}
{\sc \au{Layec, Y.} \& \au{Layec-Raphalen, M.~N.}} \yr{1983}  \at{Instability
  of dilute poly (ethylene-oxide) solutions}.  \jt{J. Phys. Lett.}
  \bvol{44}~(3),  \pg{121--128}.

\bibitem[Lecoanet {\em et~al.\/}(2016)Lecoanet, McCourt, Quataert, Burns,
  Vasil, Oishi, Brown, Stone \& O'Leary]{lecoanet2016validated}
{\sc \au{Lecoanet, D.}, \au{McCourt, M.}, \au{Quataert, E.}, \au{Burns, K.~J.},
  \au{Vasil, G.~M.}, \au{Oishi, J.~S.}, \au{Brown, B.~P.}, \au{Stone, J.~M.} \&
  \au{O'Leary, R.~M.}} \yr{2016}  \at{A validated non-linear kelvin--helmholtz
  benchmark for numerical hydrodynamics}.  \jt{MNRAS}  \bvol{455}~(4),
  \pg{4274--4288}.

\bibitem[Lellep {\em et~al.\/}(2024)Lellep, Linkmann \& Morozov]{lellep2024}
{\sc \au{Lellep, M.}, \au{Linkmann, M.} \& \au{Morozov, A.}} \yr{2024}
  \at{Purely elastic turbulence in pressure-driven channel flows}.  \jt{Proc.
  Nat. Acad. Sci.}  \bvol{121},  \pg{Ae2318851121}.

\bibitem[Li {\em et~al.\/}(2015)Li, Sureshkumar \& Khomami]{li2015simple}
{\sc \au{Li, C.~F.}, \au{Sureshkumar, R.} \& \au{Khomami, B.}} \yr{2015}
  \at{Simple framework for understanding the universality of the maximum drag
  reduction asymptote in turbulent flow of polymer solutions}.  \jt{Phys. Rev.
  E}  \bvol{92}~(4),  \pg{043014}.

\bibitem[McKinley {\em et~al.\/}(1993)McKinley, Armstrong \&
  Brown]{mckinley1993wake}
{\sc \au{McKinley, G.~H.}, \au{Armstrong, R.~C.} \& \au{Brown, R.}} \yr{1993}
  \at{The wake instability in viscoelastic flow past confined circular
  cylinders}.  \jt{Phil. Trans. R. Soc. Lond}  \bvol{344}~(1671),
  \pg{265--304}.

\bibitem[Morozov(2022)]{morozov2022coherent}
{\sc \au{Morozov, A.}} \yr{2022}  \at{Coherent structures in plane channel flow
  of dilute polymer solutions with vanishing inertia}.  \jt{Phys. Rev. Lett.}
  \bvol{129}~(1),  \pg{017801}.

\bibitem[Oliveira(2001)]{oliveira2001method}
{\sc \au{Oliveira, P.~J.}} \yr{2001}  \at{Method for time-dependent simulations
  of viscoelastic flows: vortex shedding behind cylinder}.  \jt{J. Non-Newton.
  Fluid Mech.}  \bvol{101}~(1-3),  \pg{113--137}.

\bibitem[Page {\em et~al.\/}(2020)Page, Dubief \& Kerswell]{page2020exact}
{\sc \au{Page, J.}, \au{Dubief, Y.} \& \au{Kerswell, R.~R.}} \yr{2020}
  \at{Exact traveling wave solutions in viscoelastic channel flow}.  \jt{Phys.
  Rev. Lett.}  \bvol{125}~(15),  \pg{154501}.

\bibitem[Peng {\em et~al.\/}(2023)Peng, Tang, Li, Zhang \&
  Yu]{peng2023numerical}
{\sc \au{Peng, S.}, \au{Tang, T.}, \au{Li, J.}, \au{Zhang, M.} \& \au{Yu, P.}}
  \yr{2023}  \at{Numerical study of viscoelastic upstream instability}.
  \jt{Journal of Fluid Mechanics}  \bvol{959},  \pg{A16}.

\bibitem[Pope(2000)]{Pope_turbulent2000}
{\sc \au{Pope, Stephen~B.}} \yr{2000} {\em Turbulent Flows\/}.  \publ{Cambridge
  University Press}.

\bibitem[Qin {\em et~al.\/}(2019)Qin, Salipante, Hudson \&
  Arratia]{qin2019flow}
{\sc \au{Qin, B.}, \au{Salipante, P.~F.}, \au{Hudson, S.~D.} \& \au{Arratia,
  P.~E.}} \yr{2019}  \at{Flow resistance and structures in viscoelastic channel
  flows at low re}.  \jt{Phys. Rev. Lett.}  \bvol{123}~(19),  \pg{194501}.

\bibitem[Richter {\em et~al.\/}(2010)Richter, Iaccarino \&
  Shaqfeh]{richter2010simulations}
{\sc \au{Richter, D.}, \au{Iaccarino, G.} \& \au{Shaqfeh, E.~S.~G.}} \yr{2010}
  \at{Simulations of three-dimensional viscoelastic flows past a circular
  cylinder at moderate reynolds numbers}.  \jt{J. Fluid Mech.}  \bvol{651},
  \pg{415--442}.

\bibitem[Samanta {\em et~al.\/}(2013)Samanta, Dubief, Holzner, Sch{\"a}fer,
  Morozov, Wagner \& Hof]{samanta2013elasto}
{\sc \au{Samanta, D.}, \au{Dubief, Y.}, \au{Holzner, M.}, \au{Sch{\"a}fer, C.},
  \au{Morozov, A.~N.}, \au{Wagner, C.} \& \au{Hof, B.}} \yr{2013}
  \at{Elasto-inertial turbulence}.  \jt{Proc. Natl. Acad. Sci. U. S. A.}
  \bvol{110}~(26),  \pg{10557--10562}.

\bibitem[Schneider(2015)]{schneider2015immersed}
{\sc \au{Schneider, K.}} \yr{2015}  \at{Immersed boundary methods for numerical
  simulation of confined fluid and plasma turbulence in complex geometries: a
  review}.  \jt{J. Plasma Phys.}  \bvol{81}~(6),  \pg{435810601}.

\bibitem[Shaqfeh(1996)]{shaqfeh1996purely}
{\sc \au{Shaqfeh, E.~S.~G.}} \yr{1996}  \at{Purely elastic instabilities in
  viscometric flows} ~(28),  \pg{129--185}.

\bibitem[Shekar {\em et~al.\/}(2021)Shekar, McMullen, McKeon \&
  Graham]{shekar2021}
{\sc \au{Shekar, Ashwin}, \au{McMullen, Ryan~M}, \au{McKeon, Beverley~J} \&
  \au{Graham, Michael~D}} \yr{2021}  \at{Tollmien-{S}chlichting route to
  elastoinertial turbulence in channel flow}.  \jt{Phys. Rev. Fluids}
  \bvol{6}~(9),  \pg{093301}.

\bibitem[Shekar {\em et~al.\/}(2019)Shekar, McMullen, Wang, McKeon \&
  Graham]{shekar2019critical}
{\sc \au{Shekar, A.}, \au{McMullen, R.~M.}, \au{Wang, S.~N.}, \au{McKeon,
  B.~J.} \& \au{Graham, M.~D.}} \yr{2019}  \at{Critical-layer structures and
  mechanisms in elastoinertial turbulence}.  \jt{Phys. Rev. Lett.}
  \bvol{122}~(12),  \pg{124503}.

\bibitem[Sid {\em et~al.\/}(2018)Sid, Terrapon \& Dubief]{sid2018two}
{\sc \au{Sid, S.}, \au{Terrapon, V.~E.} \& \au{Dubief, Y.}} \yr{2018}
  \at{Two-dimensional dynamics of elasto-inertial turbulence and its role in
  polymer drag reduction}  \bvol{3}~(1),  \pg{011301}.

\bibitem[Sorbie(2013)]{sorbie2013polymer}
{\sc \au{Sorbie, K.~S.}} \yr{2013} {\em Polymer-improved oil recovery\/}.
  \publ{Springer Science \& Business Media}.

\bibitem[Steinberg(2021)]{steinberg2021elastic}
{\sc \au{Steinberg, V.}} \yr{2021}  \at{Elastic turbulence: An experimental
  view on inertialess random flow}  \bvol{53},  \pg{27--58}.

\bibitem[Sureshkumar {\em et~al.\/}(1997)Sureshkumar, Beris \&
  Handler]{sureshkumar1997direct}
{\sc \au{Sureshkumar, R.}, \au{Beris, A.~N.} \& \au{Handler, R.~A.}} \yr{1997}
  \at{Direct numerical simulation of the turbulent channel flow of a polymer
  solution}.  \jt{Physics of Fluids}  \bvol{9}~(3),  \pg{743--755}.

\bibitem[Talwar \& Khomami(1995)]{talwar1995flow}
{\sc \au{Talwar, K.~K.} \& \au{Khomami, B.}} \yr{1995}  \at{Flow of
  viscoelastic fluids past periodic square arrays of cylinders: inertial and
  shear thinning viscosity and elasticity effects}.  \jt{J. Non-Newton. Fluid
  Mech.}  \bvol{57}~(2-3),  \pg{177--202}.

\bibitem[Toms(1948)]{Toms_P1INTCRHEOL1948}
{\sc \au{Toms, B.~A.}} \yr{1948} Some observations on the flow of linear
  polymer solutions through straight tubes at large {R}eynolds numbers.  \bt{In
  {\em Proc. 1st Int'l. Congress on Rheology\/}}, ,  \vol{vol.~2},  \pg{pp.
  135--141}.  \publ{Amsterdam: North-Holland}.

\bibitem[Varshney \& Steinberg(2018)]{varshney2018drag}
{\sc \au{Varshney, A.} \& \au{Steinberg, V.}} \yr{2018}  \at{Drag enhancement
  and drag reduction in viscoelastic flow}.  \jt{Phys. Rev. Fluids}
  \bvol{3}~(10),  \pg{103302}.

\bibitem[Wei {\em et~al.\/}(2014)Wei, Romero-Zer{\'o}n \& Rodrigue]{wei2014oil}
{\sc \au{Wei, B.}, \au{Romero-Zer{\'o}n, L.} \& \au{Rodrigue, D.}} \yr{2014}
  \at{Oil displacement mechanisms of viscoelastic polymers in enhanced oil
  recovery (eor): a review}.  \jt{J. Pet. Explor. Prod. Technol.}  \bvol{4},
  \pg{113--121}.

\bibitem[White {\em et~al.\/}(2012)White, Dubief \& Klewicki]{white2012re}
{\sc \au{White, C.~M.}, \au{Dubief, Y.} \& \au{Klewicki, J.}} \yr{2012}
  \at{Re-examining the logarithmic dependence of the mean velocity distribution
  in polymer drag reduced wall-bounded flow}.  \jt{Phys. Fluids}
  \bvol{24}~(2),  \pg{021701}.

\bibitem[Xi(2019)]{xi2019turbulent}
{\sc \au{Xi, L.}} \yr{2019}  \at{Turbulent drag reduction by polymer additives:
  Fundamentals and recent advances}.  \jt{Phys. Fluids}  \bvol{31}~(12).

\bibitem[Xi \& Graham(2010)]{xi2010turbulent}
{\sc \au{Xi, L.} \& \au{Graham, M.~D.}} \yr{2010}  \at{Turbulent drag reduction
  and multistage transitions in viscoelastic minimal flow units}.  \jt{J. Fluid
  Mech.}  \bvol{647},  \pg{421--452}.

\bibitem[Xi \& Graham(2012)]{xi2012intermittent}
{\sc \au{Xi, L.} \& \au{Graham, M.~D.}} \yr{2012}  \at{Intermittent dynamics of
  turbulence hibernation in newtonian and viscoelastic minimal channel flows}.
  \jt{J. Fluid Mech.}  \bvol{693},  \pg{433--472}.

\bibitem[Xiong {\em et~al.\/}(2017)Xiong, Bruneau \& Yang]{xiong2017numerical}
{\sc \au{Xiong, Y.~L.}, \au{Bruneau, C.} \& \au{Yang, D.}} \yr{2017}
  \at{Numerical study on viscoelastic fluid flow past a rigid body}.  \jt{Appl.
  Math. Model.}  \bvol{42},  \pg{188--208}.

\bibitem[Zhu {\em et~al.\/}(2024)Zhu, Atoufi, Lefauve, Kerswell \&
  Linden]{zhu2024LSA}
{\sc \au{Zhu, L.}, \au{Atoufi, A.}, \au{Lefauve, A.}, \au{Kerswell, R.~R.} \&
  \au{Linden, P.~F.}} \yr{2024}  \at{Long-wave instabilities of sloping
  stratified exchange flows}.  \jt{J. Fluid Mech.}  \bvol{983},  \pg{A12}.

\bibitem[Zhu {\em et~al.\/}(2019)Zhu, Bai, Krushelnycky \&
  Xi]{Zhu_Xi_JNNFM2019}
{\sc \au{Zhu, L.}, \au{Bai, X.}, \au{Krushelnycky, E.} \& \au{Xi, L.}}
  \yr{2019}  \at{{Transient dynamics of turbulence growth and bursting: effects
  of drag-reducing polymers}}.  \jt{J. Non-Newton. Fluid Mech.}  \bvol{266},
  \pg{127--142}.

\bibitem[Zhu \& Xi(2020)]{Zhu_XiJNNFM2020}
{\sc \au{Zhu, L.} \& \au{Xi, L.}} \yr{2020}  \at{Inertia-driven and
  elastoinertial viscoelastic turbulent channel flow simulated with a hybrid
  pseudo-spectral/finite-difference numerical scheme}.  \jt{J. Non-Newton.
  Fluid Mech.}  \bvol{286},  \pg{104410}.

\bibitem[Zhu \& Xi(2021)]{Zhu_XiPRFluids2021}
{\sc \au{Zhu, L.} \& \au{Xi, L.}} \yr{2021}  \at{Nonasymptotic elastoinertial
  turbulence for asymptotic drag reduction}.  \jt{Phys. Rev. Fluids}
  \bvol{6}~(1),  \pg{014601}.

\bibitem[Zhu \& Xi(2022)]{zhu2022direct}
{\sc \au{Zhu, L.} \& \au{Xi, L.}} \yr{2022}  \at{Direct transition to
  elastoinertial turbulence from a linear instability in channel flow}.
  \jt{arXiv preprint arXiv:2211.09366} .

\end{thebibliography}

\end{document}